\documentclass[twocolumn]{aastex631}

\usepackage{orcidlink}
\usepackage{verbatim}
\usepackage{enumerate}
\usepackage{graphicx}
\usepackage{gensymb}
\usepackage{multirow}
\usepackage{makecell}
\usepackage{ulem}

\usepackage{natbib}
\usepackage{amsmath,amsfonts,amssymb,mathrsfs,graphicx,color}
\usepackage{xcolor}
\usepackage{slashed}
\usepackage[utf8]{inputenc}
\usepackage{float}

\newcommand{\Fig}{Fig.~}
\newcommand{\Sec}{Sec.~}
\newcommand{\Tab}{Tab.~}
\newcommand{\App}{App.~}

\newcommand{\prince}{{\sc PriNCe}}
\newcommand{\xmax}[0]{X_\text{max}}

\newcommand{\TA}{\mathrm{TA}}
\newcommand{\TAO}{\mathrm{TA\phantom{O}}}
\newcommand{\PAO}{\mathrm{PAO}}
\newcommand{\cosmo}{\mathrm{cosmo}}
\newcommand{\local}{\mathrm{local}}

\newcommand{\Rmaxlocal}{$R^\mathrm{max}_\local$}
\newcommand{\Emaxlocal}{$E^\mathrm{max}_\local$}
\newcommand{\Rmaxcosmo}{$R^\mathrm{max}_\cosmo$}
\newcommand{\gammalocal}{$\gamma_\local$}
\newcommand{\gammacosmo}{$\gamma_\cosmo$}
\newcommand{\lumlocal}{$L^\mathrm{CR}_\local$}
\newcommand{\lumcosmo}{$\mathcal{L^\mathrm{CR}_\cosmo}$}
\newcommand{\mcosmo}{$m_\cosmo$}
\newcommand{\Dlocal}{$D_\local$}

\newcommand{\deltaEPAO}{$\delta_E^\PAO$}
\newcommand{\deltaETA}{$\delta_E^\TA$}

\newcommand{\deltaMeanXmaxPAO}{$\delta^\PAO_{\langle X_\mathrm{max}\rangle}$}
\newcommand{\deltaMeanXmaxTA}{$\delta^\TAO_{\langle X_\mathrm{max}\rangle}$}

\newcommand{\deltaSigmaXmaxPAO}{$\delta^\PAO_{\sigma(X_\mathrm{max})}$}
\newcommand{\deltaSigmaXmaxTA}{$\delta^\TAO_{\sigma(X_\mathrm{max})}$}

\begin{document}

\title{Differences Between the Pierre Auger Observatory and Telescope Array Spectra: Systematic Effects or Indication of a Local Source of Ultra-High-Energy Cosmic Rays?}

\author{Pavlo Plotko
\orcidlink{0000-0001-6975-5186}}
\affiliation{Deutsches Elektronen-Synchrotron DESY, Platanenallee 6, 15738 Zeuthen, Germany} 

\author{Arjen van Vliet
\orcidlink{0000-0003-2827-3361}}
\affiliation{Department of Physics, Khalifa University, P.O. Box 127788, Abu Dhabi, United Arab Emirates}
\affiliation{Deutsches Elektronen-Synchrotron DESY, Platanenallee 6, 15738 Zeuthen, Germany}

\author{Xavier Rodrigues
\orcidlink{0000-0001-9001-3937}}
\affiliation{European Southern Observatory, Karl-Schwarzschild-Straße 2, 85748 Garching bei München, Germany}
\affiliation{Astronomical Institute, Fakultät für Physik und Astronomie, Ruhr-Universität Bochum, 44780 Bochum, Germany}
\affiliation{Deutsches Elektronen-Synchrotron DESY, Platanenallee 6, 15738 Zeuthen, Germany}

\author{Walter Winter
\orcidlink{0000-0001-7062-0289}}
\affiliation{Deutsches Elektronen-Synchrotron DESY, Platanenallee 6, 15738 Zeuthen, Germany}

\begin{abstract}
The Pierre Auger Observatory (PAO) and Telescope Array (TA) collaborations report significant differences in the observed energy spectra of ultra-high-energy cosmic rays (UHECRs) above 30~EeV. In this work we present a joint fit of TA and PAO data using the rigidity-dependent maximum energy model, including a full marginalization over all relevant parameters. We test two possible scenarios to explain these differences. One is that they are due to complex energy-dependent experimental systematics; the other is the presence of a local astrophysical source in the Northern Hemisphere, which is only visible by the TA experiment. We show that the astrophysical and systematic scenarios improve the explanation of the data equally well, compared to the scenario where both experiments observe the same UHECR flux from a cosmological source distribution and have energy-independent systematics. We test different mass compositions emitted from the local source and conclude that the data are best described by a source lying at a distance below 26~Mpc that emits cosmic rays dominated by the silicon mass group. We also discuss possible source candidates, and the possible role of the putative local UHECR source in the observed TA anisotropy and in the differences in TA spectral data from different declination bands.   
\end{abstract}


\section{Introduction}
\label{sec:intro}

Ultra-high-energy cosmic rays (UHECRs) are the most energetic particles detected to date. These atomic nuclei with energies above $10^{18}$~eV are measured with increasing precision by the Pierre Auger Observatory~\citep[henceforth PAO,][]{PierreAuger:2015eyc}, located in Argentina, and the Telescope Array~\citep[TA,][]{TelescopeArray:2012uws,Tokuno:2012mi}, located in the state of Utah in the USA.
Both PAO and TA employ a hybrid detection technique to detect the extensive air showers triggered in the atmosphere by the UHECRs: a surface detector array measures the charged secondaries that reach the ground level, while fluorescence detector stations measure the development of the air showers in the atmosphere. Located in the Southern Hemisphere, PAO observes the sky below a declination of 24.8$^\circ$~\citep{PhysRevD.102.062005}, while TA, located in the Northern Hemisphere, observes the sky above -16.0$^\circ$~\citep{Ivanov:20198M}. There are therefore large portions of the Northern and Southern Hemispheres that are observed exclusively by TA and PAO, respectively, but there is also a common declination band, $-16.0^\circ<\delta<24.8^\circ$, where the sky is observed by both experiments.

\begin{figure*}[htpb!]
    \centering
    \includegraphics[width=\textwidth]{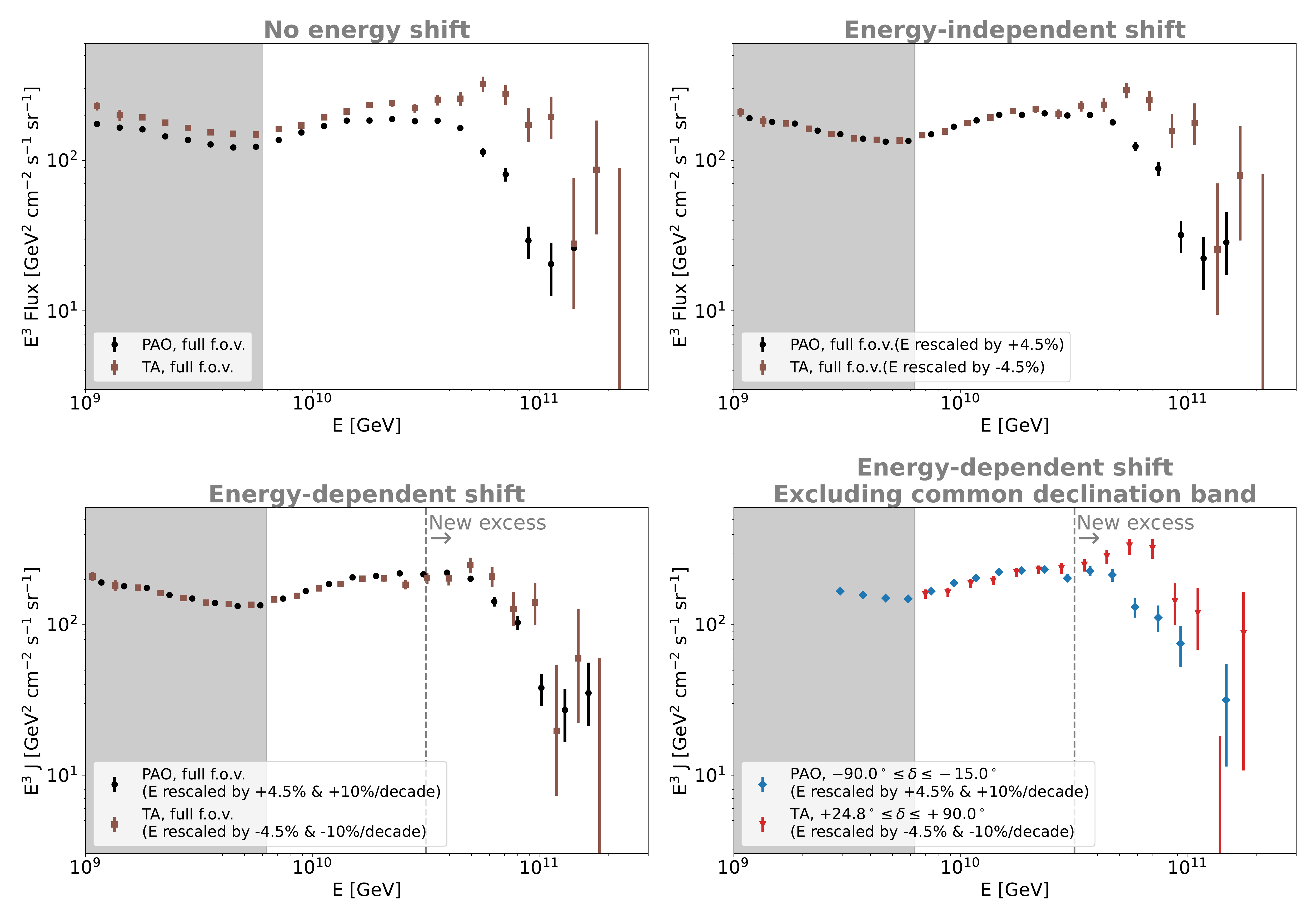}
    \caption{The black and brown data points represent the energy spectrum of the UHECRs as measured by PAO (black circles) and TA (brown squares) in the full fields of view (f.o.v.) of either observatory. In the upper left plot we see the data in the native energy scale of either experiment; in the upper right plot, those energy scales have been shifted by the amounts specified in the legend, which are within the experiments' systematics. With this energy-independent shift, the spectra are compatible at energies below 30~EeV. In the bottom left plot, the data are shifted by an energy-dependent shift (see~\citet{Tsunesada:2021qO}). The differences in the spectrum at energy above 30~EeV disappear; however, the PAO fluxes become higher than from TA at $\sim$10-30~EeV. In the bottom right plot, the same energy-dependent shift is applied to the PAO spectrum from the Southern Hemisphere (blue diamonds) and the TA spectrum from the Northern Hemisphere (red triangles). We can see that a discrepancy above 30~EeV is still present even with this energy-dependent shift.} Data adopted from the analyses by~\citet{Tsunesada:2021qO,Deligny:2019SC}.
    \label{fig:comparison}
\end{figure*}

As quantified recently by a working group from PAO and TA~\citep{Ivanov:2017wH, Deligny:2019SC, Tsunesada:2021qO}, there are differences in the energy spectrum of the UHECRs measured by both experiments, as shown in \Fig\ref{fig:comparison}, adopted from that report. In the upper left plot we see that using the energy scales native to both experiments there is a difference in the overall flux normalization as well as in the spectral shape at energies above $10^{19.5}$~eV, or about 30~EeV. Although this plot shows data from the full sky, discrepancies are also present when comparing only the common declination band of the two experiments, albeit less significantly due to higher statistical uncertainties.

The total systematic uncertainty in the energy scale of PAO is estimated at 14\%~\citep{PhysRevD.102.062005}, while for TA it is 21\%~\citep{Ivanov:20198M}. Shifting the energy scale of the experiments within these uncertainty ranges leads to a change in both the shape and normalization of the UHECR spectra, since they are plotted here as $E^3J(E)$. As shown by~\citet{Deligny:2019SC}, shifting the energy scales of the experiments by a constant value (which we will refer to henceforth as the energy-independent shift) can mitigate the spectral differences below 30~EeV, as we can see in the upper right plot of \Fig\ref{fig:comparison}. However, above that energy, the spectra become again discrepant, with TA data showing an excess in flux compared to PAO. 

For this discrepancy to be explained solely by systematic effects, energy-dependent shifts must be introduced in the energy scales of both experiments, as shown in the bottom left plot of \Fig\ref{fig:comparison}. \citet{Tsunesada:2021qO} have recently quantified the expressions for these shifts that lead to the best agreement between the two data sets in the common declination band of both experiments. At the same time, while the PAO spectrum is independent of declination, the TA spectrum seems to present different spectral features when considering events from the Northern Hemisphere compared to the common declination band (see also~\citet{Abbasi:2018ygn}). In the bottom right plot of \Fig\ref{fig:comparison} we can see clear differences between the spectra when we include the prescribed energy-dependent shift but exclude data from the common declination band of the two experiments. We can see that above $\sim$40~EeV the TA flux is higher than the one from the PAO, indicating an excess just like with the simpler assumption of an energy-independent shift, which indicates that the same energy-dependent shift obtained for the common declination band may not be sufficient to explain the differences in the entire sky~\citep{Ivanov:2017wH, Deligny:2019SC}. This suggests that either the experiment systematics are declination-dependent~\citep[which seems to be disfavored,][]{Tsunesada:2021qO}, or this effect may be astrophysical, a hypothesis that will also be tested in the present work.

The existence of a nearby UHECR source or group of sources is supported by previous analyses of the arrival directions of the TA cosmic rays. For example, using five years of data,~\citet{Abbasi_2014} reported an intermediate-scale anisotropy in the arrival directions of UHECRs above 57~EeV, with a 3.4$\sigma$ significance. More recently, an update on that analysis with twice the exposure time has revealed that this hotspot is now significant at the 3.5$\sigma$ level~\citep{Abbasi:2021arx}. In the meantime, TA has also confirmed a new excess of events with energies above 25~EeV from a different direction ~\citep{Abbasi:2021arx}. \citet{Globus:2016gvy} have suggested that a nearby UHECR source is necessary to find a consistent explanation for both the TA spectrum and anisotropy. Several astrophysical models predict that nearby sources, within 100 Mpc, could describe the PAO spectrum at the highest energies \citep{Wibig2007,Taylor2011, Mollerach2019, Lang2020}.

In this study, we investigate the hypothesis that a local source in the Northern Hemisphere is at least partially responsible for the discrepancy between the PAO and TA spectra. Using a cutting-edge numerical model, we simulate the propagation of UHECRs from a hypothetical local source observed only by TA, as well as a cosmological distribution of sources observed by the two experiments. We then perform a joint fit of the results to both TA and PAO data, considering both spectrum and composition observables and the relevant multi-parameter correlations. We then use the best-fit results to constrain the source properties. 

The paper is structured as follows: in \Sec\ref{sec:methods} we briefly describe the numerical method and specify the fitting procedure used to constrain the model parameters. In \Sec\ref{sec:results} we report the results of the fit and discuss their interpretation. In \Sec\ref{sec:conclusion} we summarize our conclusions.

\section{Methods}
\label{sec:methods}
\subsection{Source parameterization}
\label{subse:source-prop_model}

We first simulate UHECR emission from a cosmological population of UHECR sources. The sources are considered to be homogeneously distributed and have an emissivity in cosmic rays \lumcosmo, defined as the emitted cosmic-ray luminosity per cosmic volume, and obtained by integrating the emission spectrum above $10^9$~GeV. Each emitted isotope of mass $A$ contributes a fraction $f_A$ of that emissivity. The set of all $f_A$ values defines the emitted composition, which we assume to also be the same for all sources. The spectrum of each element $A$ emitted by the cosmological source distribution, $J_{A}^\cosmo$, can be parameterized as~\citep{Aab:2017,Batista:2019,Heinze:2019}:
\begin{equation} \label{eq:emissivity}
\begin{split}
J_A^\cosmo{}(E,z) =& J_0^\cosmo~f_A\,n(z,\text{\mcosmo})  \\
     &\left(\frac{E}{10^9~\text{GeV}} \right)^{-\text{\gammacosmo}} f_\mathrm{cut}(E),
\end{split}
\end{equation}
where $n=(1 + z)^\text{\mcosmo}$ is the cosmological source density (the index \mcosmo\ characterizes the evolution of the source density with redshift $z$), \gammacosmo~is the spectral index of the emitted cosmic rays, $J_0^\cosmo$ is the normalization of the spectra that corresponds to the total emissivity, and the factor $f_\mathrm{cut}$ introduces an exponential cutoff at the energy corresponding to the maximum rigidity $\text{\Rmaxcosmo} = E_\mathrm{max}/Z_\text{A}$:

\begin{align}
    f_\text{cut}(E) = \begin{cases}
        1 
        &, E < Z_\text{A} \text{\Rmaxcosmo}, \\
        \exp\left(1 - \frac{E}{Z_\text{A} \text{\Rmaxcosmo}} \right)
        &, E > Z_\text{A} \text{\Rmaxcosmo}.
    \end{cases}
\end{align}
The maximum rigidity \Rmaxcosmo~of all emitted isotopes is the same, as is typical of  astrophysical sources optically thin to nuclear disintegration~\citep{Kotera_2015,Rodrigues:2017fmu,Biehl:2017zlw}. The fraction of injection elements ($f_A$) is defined at a fixed energy of $10^9$ GeV, relative to a total normalization. We provide integrated fractions of the energy density ($I^9_A$), since that definition is less sensitive to the arbitrary choice of the energy value:
\begin{align}
\label{eq:integral_fractions}
I^9_A = \frac{\int_{10^{9}~\text{GeV}}^\infty J_A(E) E \text{d}E}{\sum_A \int_{10^{9}~\text{GeV}}^\infty J_A(E) E \text{d}E}.
\end{align}

Regarding the injection composition (i.e. the chemical composition of the cosmic rays as they are emitted by the sources), we consider a combination of five elements, each representative of a distinct mass group up to iron-56: $^1$H, $^4$He, $^{14}$N, $^{28}$Si and $^{56}$Fe. In terms of the propagation simulation we start with a combination of these five isotopes at the source (whose fractions are given by $I^9_A$ as discussed above). As they interact with the cosmic photon backgrounds, they then produce nuclear cascades of hundreds of possible secondary isotopes with intermediate masses, all of which are taken into account.

To test the local source hypothesis we will assume additionally the existence of a single local source in the Northern Hemisphere that can be observed by TA but not by PAO. For simplicity, we only explore scenarios where the local source emits a single mass group from the five listed above, which is done by propagating the respective representative isotope. Therefore, the emission from this source can be fully characterized by a maximum rigidity \Rmaxlocal, power-law index \gammalocal, emission luminosity \lumlocal, and one emitted cosmic-ray mass group. These parameters will be varied independently from those describing the cosmological source distribution. Finally, the comoving distance to the local source\footnote{We will generally refer to the comoving distance to the local source, while keeping in mind that the cosmic rays generally travel a longer distance due to magnetic field deflections, which cannot be included in a one-dimensional simulation. This distinction becomes more relevant the larger the value of \Dlocal.} will affect the UHECR spectrum observed at Earth, in a similar manner to how the evolution parameter \mcosmo~affects the contribution from the cosmological source distribution.

Overall, the cosmological source distribution can be fully characterized by eight parameters $\boldsymbol\lambda_\cosmo$ and the local source by five parameters $\boldsymbol\lambda_\local$:
\begin{align}
    \begin{split}
        \boldsymbol\lambda_\cosmo \,=\, 
        &( \text{\gammacosmo}, \text{\Rmaxcosmo},  \text{\mcosmo},\text{\lumcosmo},I^{9,\cosmo}_{\text{A}}),\\
        \boldsymbol\lambda_\local \,=\, 
        &( \text{\gammalocal}, \text{\Rmaxlocal}, \text{\Dlocal}, \text{\lumlocal}, A_\local).
        \label{eq:parameter_sets}
    \end{split}
\end{align}

\subsection{Propagation model}
\label{subsec:model}

Once we characterize the emitted UHECR spectra from the cosmological source distribution and the local source with two sets of parameters $\boldsymbol\lambda_\cosmo$ and $\boldsymbol\lambda_\local$ (Eq.~\ref{eq:parameter_sets}), we then inject those cosmic rays into a numerical simulation to calculate their interactions as they propagate toward Earth. For this, we use the open-source software \prince~~\citep{Heinze:2019}  which numerically solves the transport equations of the cosmic-ray spectra. We use \textsc{Talys}~\citep{Koning:2007} as the nuclear interaction model and we adopt the Extragalactic Background Light (EBL) model by~\citet{Gilmore:2012}. 

In the case of the cosmological source distribution we consider the emission from a continuous distribution of sources up to $z=1$, since a source at any higher redshift will lie outside the cosmic-ray horizon for the energy range we focus on. In each simulation we assume a certain value of the evolution parameter \mcosmo, which defines the strength of the cosmological evolution of the source distribution, as defined in Eq.~(\ref{eq:emissivity}).

We then add the contribution from the local source. In the \prince~framework, a single source lying at a comoving distance \Dlocal\ is equivalent to a source population whose evolution is described by a delta function, $n=\delta(D=\text{\Dlocal})$. Because the cosmological source distribution and the local source are independent, we simulate the propagation of the two separately. We then add the respective contributions of the propagated spectrum at Earth to obtain the prediction for the total spectrum observed by TA, while for PAO observables only the contribution from the cosmological source distribution is considered. 

After simulating the cosmic ray propagation according to the above procedure, we obtain the energy spectrum for each individual isotope at the top of the atmosphere, as well as values of $\langle \ln{A} \rangle$ and $\sigma^2_{\ln{A}}$ for each energy of the numerical grid. We then compute the mean of the distribution of the depth of the shower maximum,  $\langle \xmax \rangle$, as well as its second moment, $\sigma(\xmax)$, following the procedure by~\citet{Heinze:2019}. Throughout this work we will discuss the results for three air shower models separately: \textsc{Sibyll~2.3}c~\citep{riehn:2016hO},
\textsc{Epos-LHC}~\citep{Pierog:2015}, and \textsc{QGSJET-II-04}~\citep{Ostapchenko:2011}. 
These predictions will then be compared with data from both the TA and PAO experiments, resulting in a joint fit.

\subsection{Joint fit of PAO and TA data}
\label{subsec:fit}

We aim at describing the spectrum and composition of UHECRs above ${E_\text{min} = 6\times10^9 ~ \text{GeV}}$, originating from the entire field of view of TA and PAO. The data from PAO consists of spectrum measurements distributed over fourteen energy bins~\citep{Verzi:2019AO}, nine data points describing $\langle X_{\mathrm{max}}\rangle$, and nine more for $\sigma(X_{\mathrm{max}})$~\citep{Yushkov:2019J8}. The TA spectrum above our threshold is described by fifteen data points and one upper limit~\citep{Ivanov:20198M}, while the data on $\langle X_{\mathrm{max}}\rangle$ and $\sigma(X_{\mathrm{max}})$ consist of five data points each\footnote{Ideally, we could draw additional information by making use of data subsets from the Northern Hemisphere, Southern Hemisphere, and the common declination band seen by both experiments.}~\citep{Abbasi2018}. In addition, we include five spectrum data point from either observatory at energies above $2\times10^9 ~ \text{GeV}$ lower than $6\times10^9 ~ \text{GeV}$  in our analysis to constrain our theoretical predictions of the spectrum and ensure that we do not overshoot the data.

Regarding the composition observables, i.e. $\langle X_{\mathrm{max}}\rangle$ and $\sigma(X_{\mathrm{max}})$, we adopt the values published by the two experiments independently. As argued by~\citet{deSouza:2017ZU}, a detailed comparison should take into account the different detector acceptances and resolutions, as well as the differences between the analysis techniques of the two groups. However, the tools and data for such detailed treatment of the TA composition have not thus far been disclosed. Furthermore, as shown in \Sec\ref{sec:results} and previously by~\citet{Heinze:2019}, the three air shower models considered in this work lead to different predictions on the observed composition, which introduces an element of uncertainty that surpasses the uncertainty from these more detailed effects. We, therefore, compare the composition data at face value and leave a more precise analysis for a future work.

As described in \Sec\ref{subse:source-prop_model}, we hypothesize that these data are explained by a cosmological source distribution, characterized by eight parameters $\boldsymbol\lambda_\cosmo$, and a local source in the Northern Hemisphere, characterized by five parameters $\boldsymbol\lambda_\local$. To account for the systematic uncertainties of TA and PAO, we further introduce six nuisance parameters $\boldsymbol\delta$: $\delta_E^\TA$ and $\delta_E^\PAO$ characterize the uncertainties in the energy scales of the TA and PAO spectra, respectively  (see \Sec\ref{sec:intro}). Furthermore, we have $\delta^\TA_{\langle X_\mathrm{max}\rangle }$ and $\delta^\PAO_{\langle X_\mathrm{max}\rangle }$, which define systematic shifts in $\langle X_{\mathrm{max}}\rangle$, and $\delta^\PAO_{\sigma(X_\mathrm{max})}$ and $\delta^\TAO_{\sigma(X_\mathrm{max})}$, which define a systematic shift in $\sigma(X_{\mathrm{max}})$.

As discussed in \Sec\ref{sec:intro}, two forms have been proposed for the energy shifts \deltaEPAO and \deltaETA and we explore both these possibilities. In the first case, we assume energy-independent energy shifts, where \deltaEPAO and \deltaETA are constant for each energy bin. In the second, more complex scenario, we consider energy-dependent systematic energy shifts. To investigate this scenario, we apply a 10\% energy shift per decade to the spectrum data, while the constant part (\deltaEPAO and \deltaETA) of the shift is treated as a free parameter. In both cases, $\delta_{E}$ are given as a percentage of each energy bin. We explore the values of $\delta$ within the 1$\sigma$ uncertainty range of each experiment ($\pm$14\% for $\PAO$ and $\pm$21\% for $\TA$ as described previously).

There is no precedent in the literature for the treatment of the systematic shifts of the composition observables in a joint fit, because to date no such joint fit has been performed. We assume that $\delta_{\langle X_\mathrm{max}\rangle}$ is given as a percentage of the systematic uncertainty of each $\xmax$ data point, and likewise $\delta_{\sigma(\xmax)}$ as a percentage of the systematic uncertainty of each $\sigma(\xmax)$ data point. For the $\sigma(\xmax)$ observable, the systematic error is of the order of a few percent for the PAO data, but the TA systematic uncertainties can be as high as several tens of percent. While $\delta^\PAO_{\sigma(\xmax)}$ could be neglected due to the small values, $\delta^\TAO_{\sigma(\xmax)}$ cannot be neglected. We therefore decide to include both nuisance parameters in the joint fit to treat both data sets equally. We explore the values of $\delta$ within the 1$\sigma$ uncertainty range of each data point.

We treat the $\boldsymbol\delta$ variables as nuisance parameters, independently from each other and from the source parameters $\boldsymbol\lambda_\cosmo$ and $\boldsymbol\lambda_\local$. We search the range $\pm100\%$, which represents the $1\sigma$ boundaries of the systematic uncertainties carried by the data from either experiment. We consider only energy-independent values of $\delta_{\langle X_\mathrm{max}\rangle}=\mathrm{const.}$ and $\delta_{\sigma(\xmax)}=\mathrm{const.}$.

The goodness of fit relative to the PAO and TA data is calculated by means of a $\chi^2$ test:
\begin{align}
    \begin{split}
    \chi^2_\PAO \,=\,& 
        \chi^2_\PAO(
        \boldsymbol{\lambda}_\cosmo,\, 
        \delta^\PAO_E,\,
        \delta^\PAO_{\langle X_\mathrm{max}\rangle },\,
        \delta^\PAO_{\sigma(X_\mathrm{max})}). \\
        \chi^2_\TAO \,=\,& 
        \chi^2_\TAO(
        \boldsymbol{\lambda}_\cosmo,\, 
        \boldsymbol{\lambda}_\local,\, 
        \delta^\TAO_E,\,
        \delta^\TAO_{\langle X_\mathrm{max}\rangle},\,
        \delta^\TAO_{\sigma(X_\mathrm{max})} ),
    \end{split}
\end{align}
where we assume that the cosmological source distribution characterized by $\boldsymbol\lambda^\mathrm{cosmo}$ is observed by both experiments.

Finally, to compute the goodness of the joint fit, we combine the $\chi^2$ values from both experiments with the systematic uncertainties:\footnote{Here we follow \citet{Huber:2004ka, Huber:2007ji}, who employed similar methods to combine the data from multiple neutrino oscillation experiments.}
\begin{align}
    \chi^2_\mathrm{global}(\boldsymbol\lambda_\cosmo,&\, \boldsymbol\lambda_\local,
    \boldsymbol\delta) = \\ 
    &\chi_\PAO^{2} + \left(\frac{\delta_E^\PAO }{\sigma_{\text{E}} ^\PAO}\right)^2 \nonumber\\
    & + \left( \frac{\delta_{\langle X_\mathrm{max}\rangle} ^\PAO}{100\%}\right)^2+ \left(\frac{\delta_{\sigma(X_\mathrm{max})} ^\PAO}{100\%}\right)^2\nonumber\\
    &+\chi_\TAO^{2} + \left(\frac{\delta_E^\TAO }{\sigma_{\text{E}}^\TAO}\right)^2 \nonumber\\
    &+ \left(\frac{\delta_{\langle X_\mathrm{max}\rangle} ^\TAO}{100\%} \right)^2+ \left( \frac{\delta_{\sigma(X_\mathrm{max})} ^\TAO}{100\%}\right)^2. \nonumber
\end{align}

As we can see, this $\chi^2$ value takes into account both the energy spectrum and the composition observables with energies above $E_\mathrm{min} = 6\times10^{9} \text{ GeV}$ from both experiments. To account for energy-dependent shifts in our analysis we include only the constant part of the shift, treating it as a free parameter, while fixing the energy-dependent part to a 10\% energy shift per decade.

A simultaneous scan of all parameters $\boldsymbol\lambda_\cosmo$, $\boldsymbol\lambda_\local$ and $\boldsymbol\delta$ would be computationally expensive, so instead we divide it into two steps. First, we perform a scan of $\boldsymbol\lambda_\cosmo$ assuming only a cosmological source distribution. We consider the spectral and composition data from PAO in our entire energy range as well as TA data below 25 EeV. This allows us to constrain the source distribution parameters $\boldsymbol\lambda^\mathrm{cosmo}$ as well as the systematic variables $\boldsymbol\delta$ of both PAO and TA. We scan a three-dimensional parameter grid in \gammacosmo$\times$\Rmaxcosmo$\times$\mcosmo\ with 81$\times$61$\times$61 elements. For each set of parameter values we numerically simulate the propagation of the five different primary isotopes from the entire source population, as described previously. The result of the propagation for different values of the other parameters (\lumcosmo~and the five $\boldsymbol I_A^{9,\mathrm{cosmo}}$ parameters) are obtained by fitting the spectra and composition data from PAO and TA observations. 

We then fit the TA data in the full energy range, assuming the experiment observes (1) the cosmological source distribution, with the parameter values $\boldsymbol\lambda^\mathrm{cosmo}$ previously obtained, and (2) a local source characterized by $\boldsymbol\lambda^\mathrm{local}$, which we now optimize. For that, we scan a fine grid in \gammalocal×\Rmaxlocal×\Dlocal~with 40$\times$61$\times$80 elements. For each set of parameter values we simulate the propagation of an UHECR spectrum composed of a single isotope $A^\mathrm{local}$ from the local source to Earth. The fluxes arriving at Earth from the local source are added to those from the best-fit cosmological source distribution, before fitting the TA data.

This two-step approach does not affect the results because above 30~EeV, where the experiments differ, the overall fit is driven mainly by the PAO spectrum (due to its lower uncertainties), and therefore depends primarily on the cosmological source distribution. The local source parameters can then be searched separately in step two, in order to optimize the fit to high-energy TA data.

This method differs from that employed by~\citet{Heinze:2019} in four aspects: 1) that work considers only the PAO spectrum and composition data, while we include TA data in the same fit; 2) that work considers only a cosmological source distribution, while we consider also the presence of a local source observed by TA; 3) that work takes into account only uncertainties in the energy scale, while we include also the uncertainties on $\langle X_{\mathrm{max}}\rangle$ and $\sigma(X_\mathrm{max})$; and 4) that work does not include constraints on the predicted spectrum for energies lower than ${E_\text{min}}$, while we include the data in this energy region as upper limits to ensure that our predictions do not overshoot the observed fluxes.

Using this method, we test four different scenarios: 1) the null hypothesis, where both TA and PAO observe the same cosmological source distribution and an energy-independent shift; 2) the assumption that TA additionally observes a local source in the Northern Hemisphere, along with an energy-independent systematic energy shift; 3) the same consideration of a cosmological source distribution as in the first scenario but with an energy-dependent systematic energy shift; and 4) the presence of a local source in the Northern Hemisphere observed by TA, along with an energy-dependent systematic energy shift. To evaluate our hypothesis, we  use the Akaike Information Criterion with correction~\citep[$\mathrm{AIC}_\mathrm{c}$,][]{Akaike:74,Kenneth:2004, Buchner:2014,Rosales:2020}.

For each scenario, including the null hypothesis, we calculate the $\mathrm{AIC}_\mathrm{c}$ value using the following formula: 
\begin{align}
    \mathrm{AIC}_{\mathrm{c},i} = \chi^2_{\mathrm{global,} i} + 2k_i + \frac{2k_i^2+2k_i}{n-k_i-1}, 
\end{align}
where $n$ is the number of data points (56) and $k_i$ is the number of parameters of the corresponding model $i$. The model with the smallest $\mathrm{AIC}_\mathrm{c}$ provides the best fit to the data. However, it's important to note that the absolute value of $\mathrm{AIC}_\mathrm{c}$ is not interpretable. Instead, $\mathrm{AIC}_\mathrm{c}$ is meaningful in a relative sense, when compared to other $\mathrm{AIC}_\mathrm{c}$ values. In our analysis, we compare the $\mathrm{AIC}_\mathrm{c}$ of each scenario to the $\mathrm{AIC}_\mathrm{c}$ of the null hypothesis (AIC$_{\mathrm{c}\mathrm{, null}}$). We determine the differences ($\Delta_i$) as follows:
\begin{align}
    \Delta_{i} = \mathrm{AIC}_{\mathrm{c}\mathrm{, null}} - \mathrm{AIC}_{\mathrm{c},i}. 
\end{align}

In order to simplify the comparison between models, we convert  $\Delta_i$ to the standard deviations, representing the deviation of the observed difference in $\mathrm{AIC}_\mathrm{c}$ values from the null hypothesis. It's worth noting that $\mathrm{AIC}_\mathrm{c}$ should not be used together with the null hypothesis significance test, due to the different nature of the two analysis paradigms~\citep{Anderson2000, Mundry2021}. In our study, we use the $\mathrm{AIC}_\mathrm{c}$ only as a rough comparison of the goodness of models. We do not make any strong conclusions based on the number of standard deviations. We use the relative likelihood generalization based on $\mathrm{AIC}_\mathrm{c}$~\citep{Anderson2000}. We convert the $\Delta_i$ into $p$-values using the formula: 
\begin{align}
    p_{i} = \exp \left( -0.5 \Delta_i \right). 
\end{align}
We can then determine the corresponding number of standard deviations by employing the cumulative distribution function of a standard normal distribution.

Furthermore, for completeness, we also performed a fit of the model of a cosmological source distribution to TA data only. The results of this fit are discussed in \App\ref{app:TA} and compared to the main results of this study as well as the previous work by~\citet{Heinze:2019}. As we will show, when fitting only TA data, a cosmological source distribution provides a slightly worse fit in comparison to fitting only PAO data (best-fit $\chi^2$/d.o.f.~$=1.6$ vs.~1.3) The joint fit obtained in the main part of this paper leads to a higher value of $\chi^2$/d.o.f.~$=1.7$. As discussed in greater detail in \Sec\ref{sec:results}, this is simply due to the inclusion of data from both experiments, which does not allow for a lower chi-squared value regardless of the model being fitted.

\section{Results and Discussion}
\label{sec:results}

\begin{figure*}[htbp!]
    \centering
    \begin{tabular}{cc}
       \includegraphics[width=.45\textwidth]{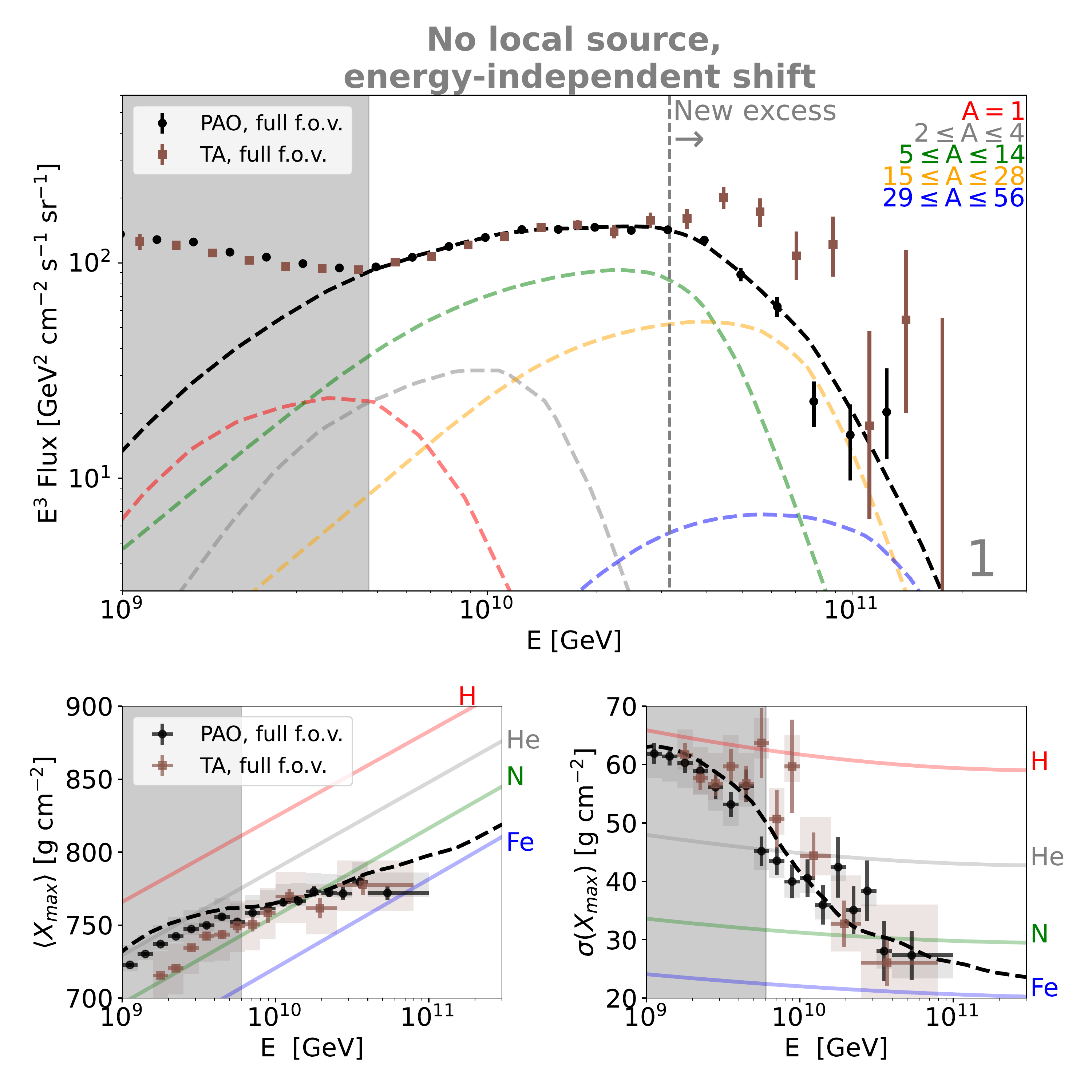} & \includegraphics[width=.45\textwidth]{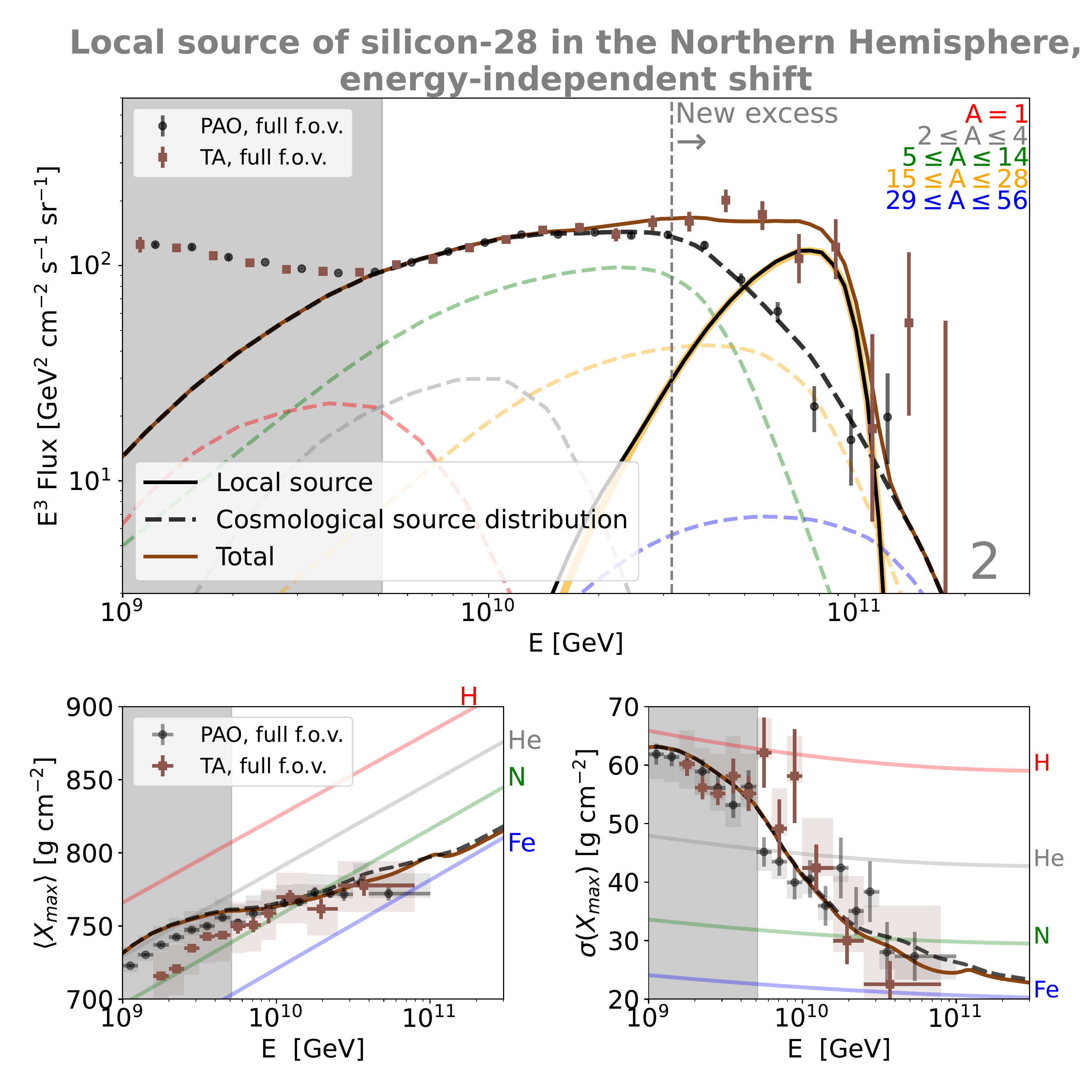} \\
        \includegraphics[width=.45\textwidth]{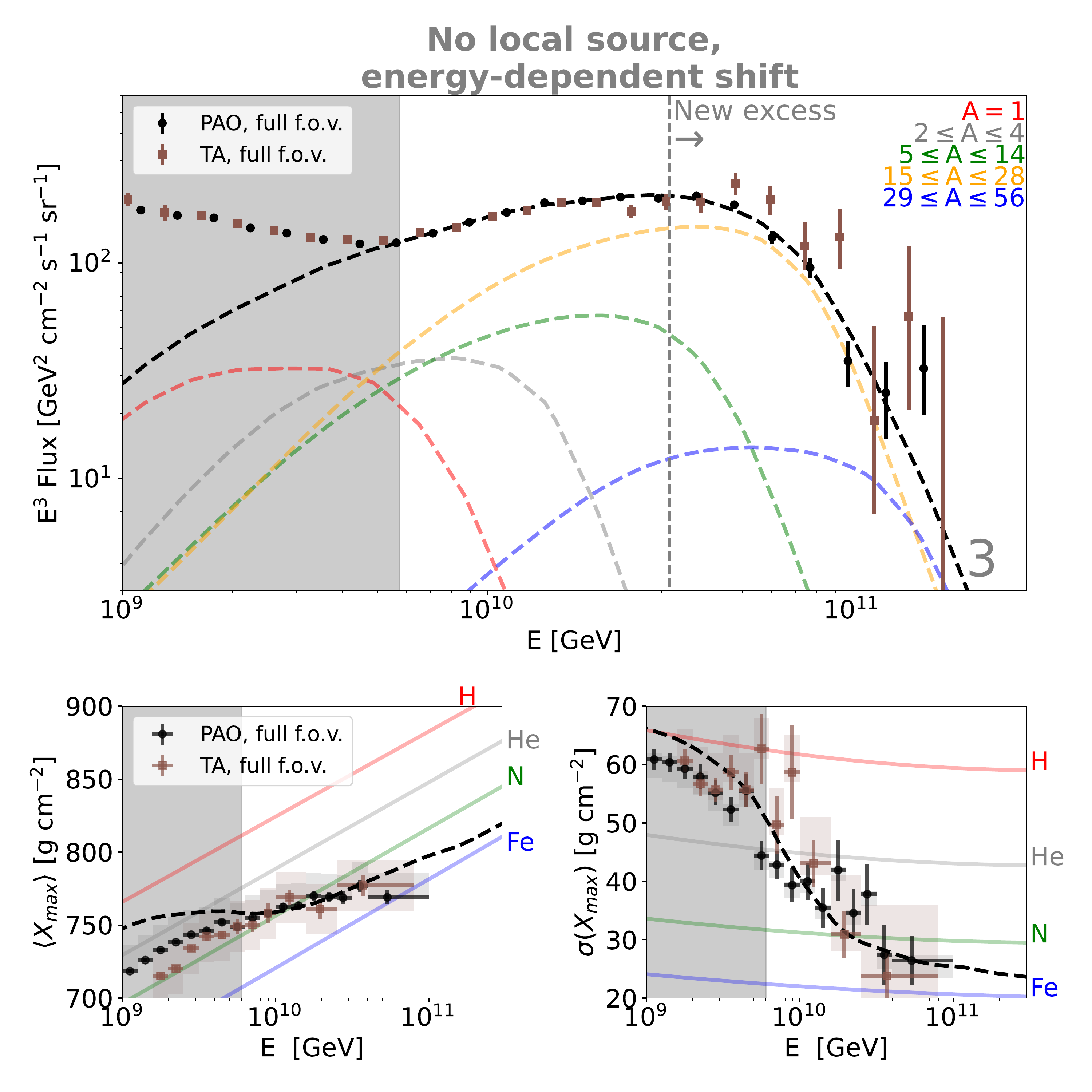}&
        \includegraphics[width=.45\textwidth]{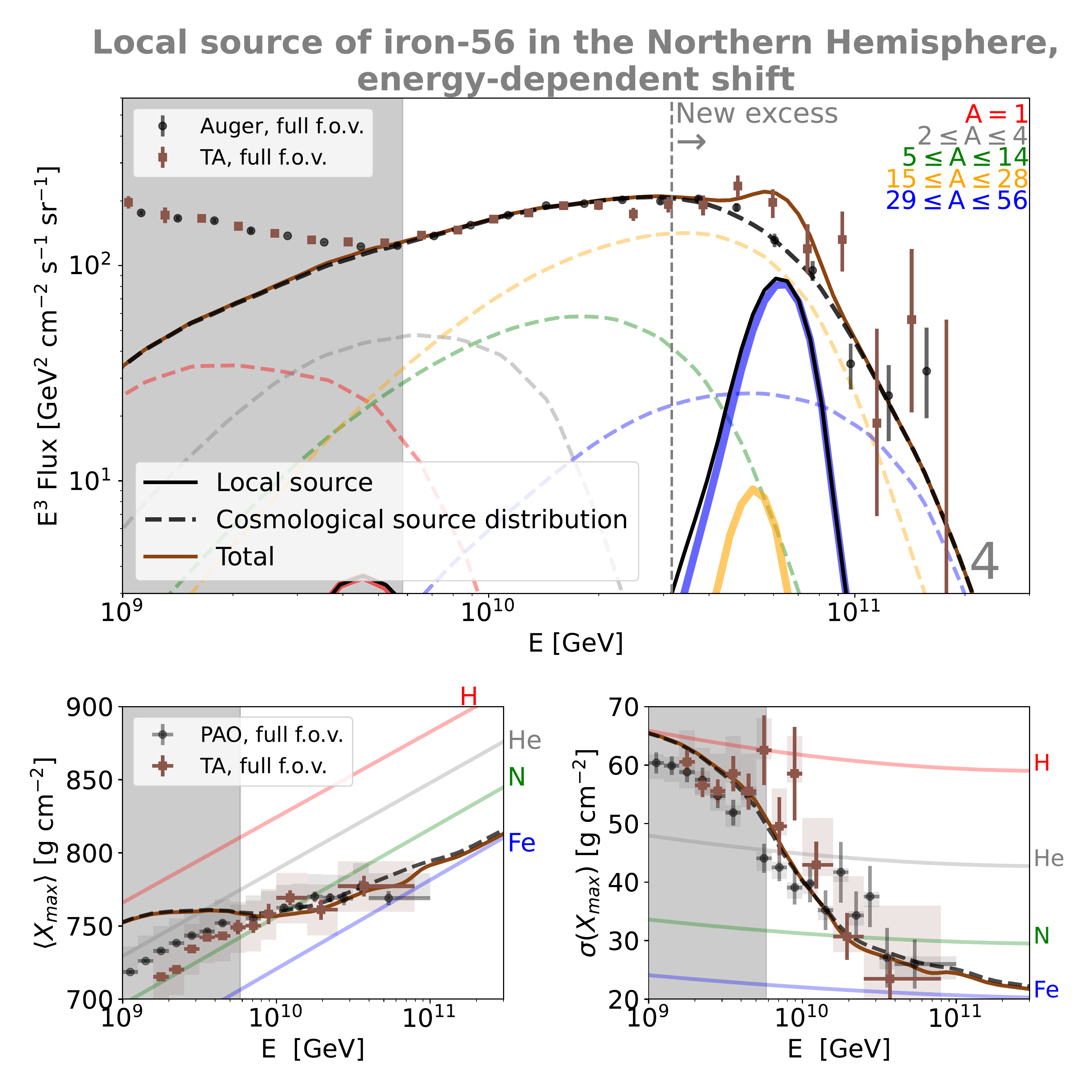}\\
    \end{tabular}
    \caption{Spectra (upper plot in each panel) and composition observables (two lower plots in each panel) resulting from a joint fit to TA and PAO data, using \textsc{Sibyll 2.3}c as the air shower model. All scenarios include a cosmological source distribution. The upper panels show the best fits considering  energy-independent shifts in the energy scales of the experiments, without a local source (left) and with the additional presence of a local source of silicon-28 in the Northern Hemisphere (right). The bottom plots show the best fits considering energy-dependent shifts in the energy scales of the experiments without a local source (left) and with the additional presence of a local source of iron-56 in the Northern Hemisphere, respectively (right). The best-fit parameter values are given in \Tab\ref{tab:main_result_parameters}. In the energy range shaded in gray, data are included only as upper limits.}
    \label{fig:main_result_predictions}
\end{figure*}

All scenarios are evaluated using the joint fit method described in \Sec\ref{sec:methods}, considering both TA and PAO data, and assuming systematic shifts, which are also optimized as part of the fit. In~\Fig\ref{fig:main_result_predictions} we show the best-fit results in all four scenarios, obtained using \textsc{Sibyll} as the air shower model. The results for \textsc{Epos-LHC} and \textsc{QGSJET} air shower models are presented in \App\ref{app:other-air-shower-models}. The cases with energy-independent shifts and those with energy-dependent shifts are represented in the upper and lower panels, respectively. Simultaneously, the left-hand side panels represent the scenarios without a local source, while the right-hand side ones represent the inclusion of a local source in the Northern Hemisphere. The best fit was found for silicon-28 and iron-56 at distances of 13.9~Mpc and 195 Mpc, for the case of an energy-independent and energy-dependent shift, respectively. The exact values and uncertainties of the best-fit parameters of the cosmological source distribution, the local source and the systematic shifts are provided in \Tab\ref{tab:main_result_parameters}. For  four  mass groups emitted by the local source that do not lead to best fits, there is a dedicated table in \App\ref{app:other_isotopes}.

In each panel of \Fig\ref{fig:main_result_predictions}, the upper plot shows the predicted cosmic-ray spectra. Focusing first on the data, we show as black points the energy-shifted PAO spectrum and as brown points the energy-shifted TA spectrum. As in the remainder of this paper, we consider data from the entire field of view of both experiments. For reference, we show as a vertical dashed line the threshold energy of 25~EeV above which the new excess was found in the TA data~\citep{Abbasi:2021arx}. In \Tab\ref{tab:main_result_parameters} we provide the best-fit values for the energy-independent and energy-dependent shifts. As expected, the relative systematic shift between the energy scales, $\delta_E^\PAO-\delta_E^\TA$, is consistent with the previous analysis by~\citet{Tsunesada:2021qO}. The absolute values of $\delta_E^\PAO$ and $\delta_E^\TA$  differ from that study because there they were chosen to be symmetric, while here they were obtained through the joint fit to our model. Like in \Fig\ref{fig:comparison}, this energy-independent shift leads to an agreement between the spectra from the two experiments up to about 30~EeV, while above that energy the TA fluxes are generally higher. The energy-dependent systematics can explain the differences at energies above that energy. Nonetheless, using this shift leads to discrepancies at energies between 20 and 30~EeV for data from the full field of view, where data are comparable with the energy-independent shift. This is because the energy-dependent shift is optimized only for the common declination band of both experiments.

The dashed black curves represent the contribution of the cosmological source distribution, which is observed by both experiments. The contributions from the different mass groups are shown in different colors. In the top right and bottom right plots we show, in addition, as a solid black curve the contribution from the local source. Because the source is relatively close by, its contribution consists almost entirely of the same mass group originally emitted by the source, as indicated by the solid yellow curve (top right) and solid blue curve (bottom right) that can be seen behind the black curves. Finally, the brown curves represent the sum of the cosmological source distribution and the local source, which is the total flux observed by TA.

The first thing to note is that the best-fit spectrum from the cosmological source distribution is similar in all cases, although in the left-hand plots the contribution from the cosmological source distribution is fitted to full data sets from both experiments and in the right-hand plot only to the full data set from PAO and TA data below 25~EeV. This is because at higher energies, where the PAO and TA spectra are discrepant, the TA data have considerably larger uncertainties, and therefore the fit of the cosmological source distribution is driven by the PAO data in all scenarios. This can also be seen by comparing the columns of \Tab\ref{tab:main_result_parameters}, where the parameters of the source population are in fact similar in all scenarios.

Comparing the best-fit results with data, we can see that the null hypothesis (upper left panel) of the cosmological source distribution with energy-independent shift alone fails to explain TA data above 30~EeV, and the overall joint fit has a high $\chi^2$ value of $110.6/54=2.0$ per degree of freedom (d.o.f.), which includes also the composition data in the bottom plots.

In contrast, models 2, 3, and 4 have lower values of $\chi^2$/d.o.f. at 1.4, 1.5 and 1.5, respectively, because the tension with TA data is explained by the additional flux from the local source, or the energy-dependent shift, or both. While in the null hypothesis the large value of $\chi^2$/d.o.f. was mainly due to the tension with the TA spectrum data at high energies, in these cases, the fit cannot be further improved due to the low uncertainties of the PAO and TA spectra at energies below $\sim20$~EeV. 

Even by using the energy-dependent shift, the data from both experiments cannot be brought to a more precise agreement at these low energies, which necessarily limits the quality of any joint fit. This is in contrast with the PAO-only fit by~\citet{Heinze:2019}, who obtained a lower $\chi^2$/d.o.f. of 1.3.

\begin{figure*}[htp!]
    \centering      
        \begin{tabular}{cc}
       \includegraphics[width=.45\textwidth]{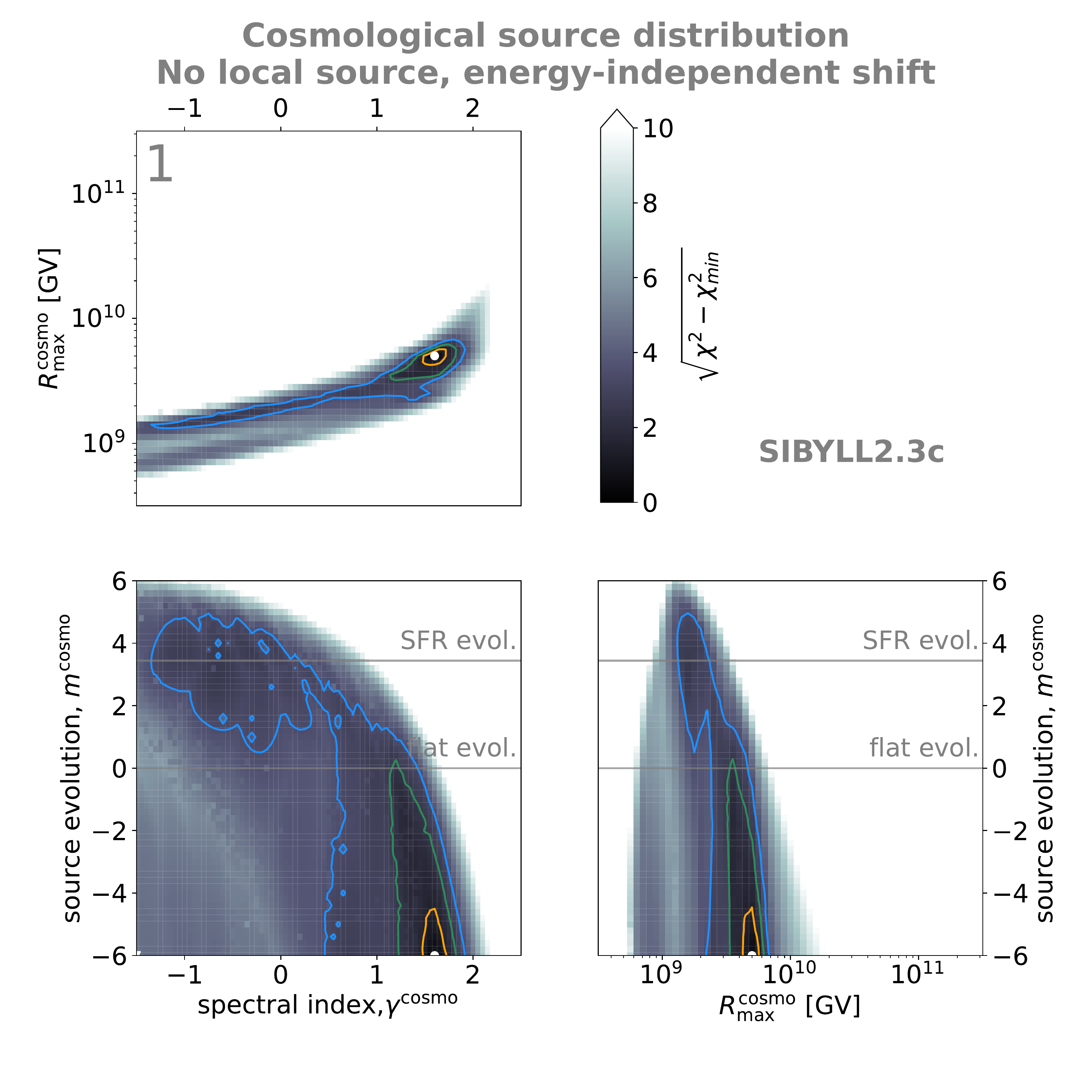} & \includegraphics[width=.45\textwidth]{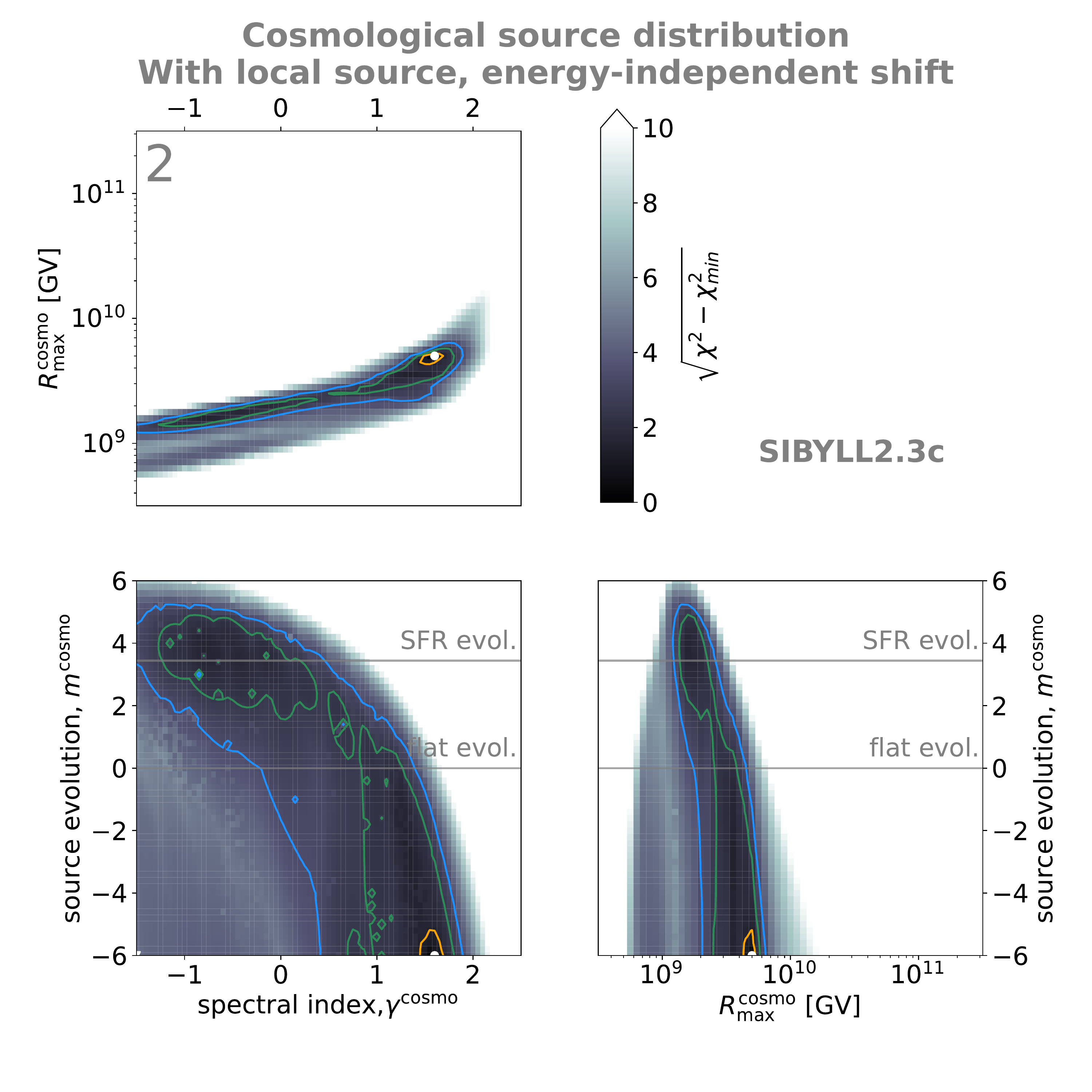} \\
        \includegraphics[width=.45\textwidth]{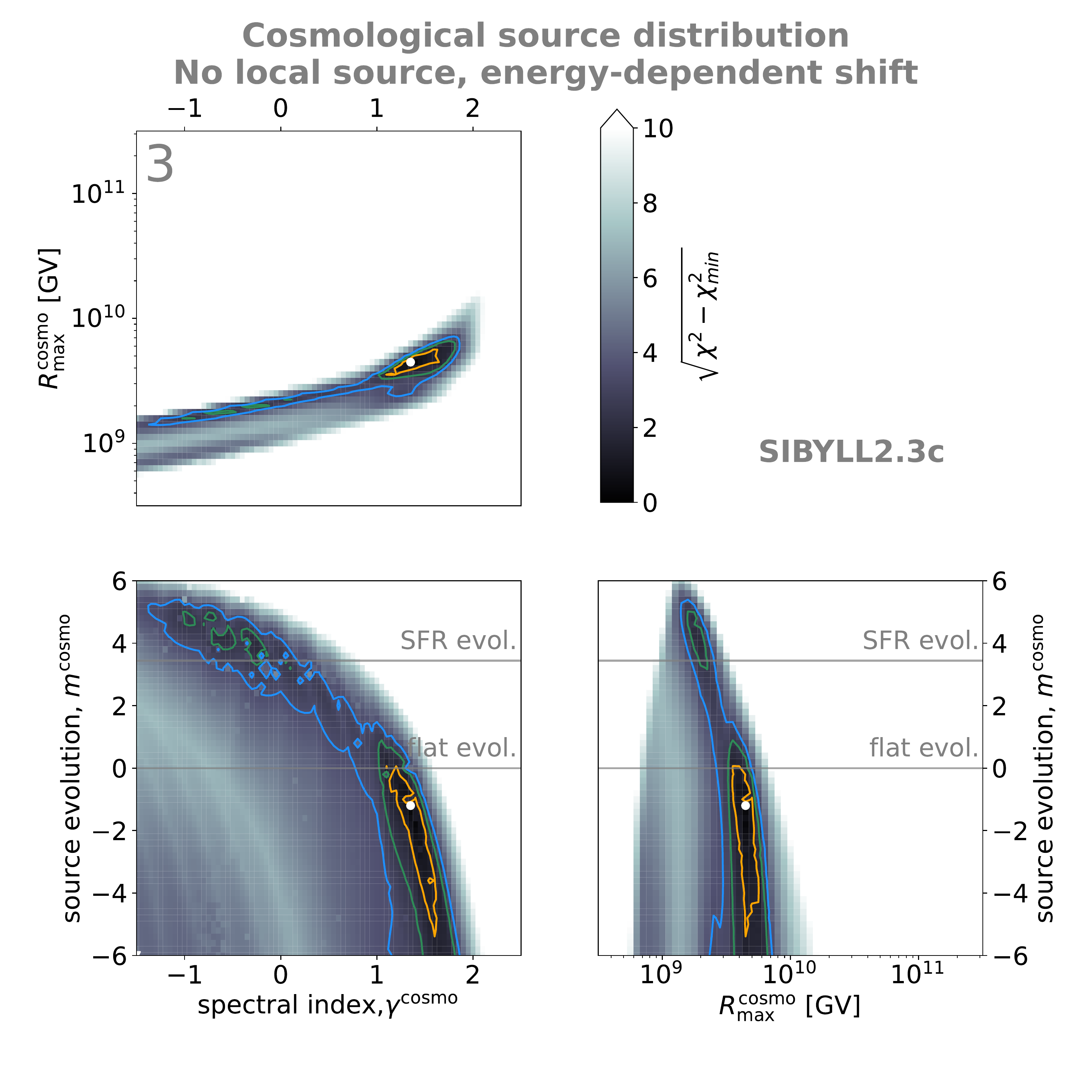}&
        \includegraphics[width=.45\textwidth]{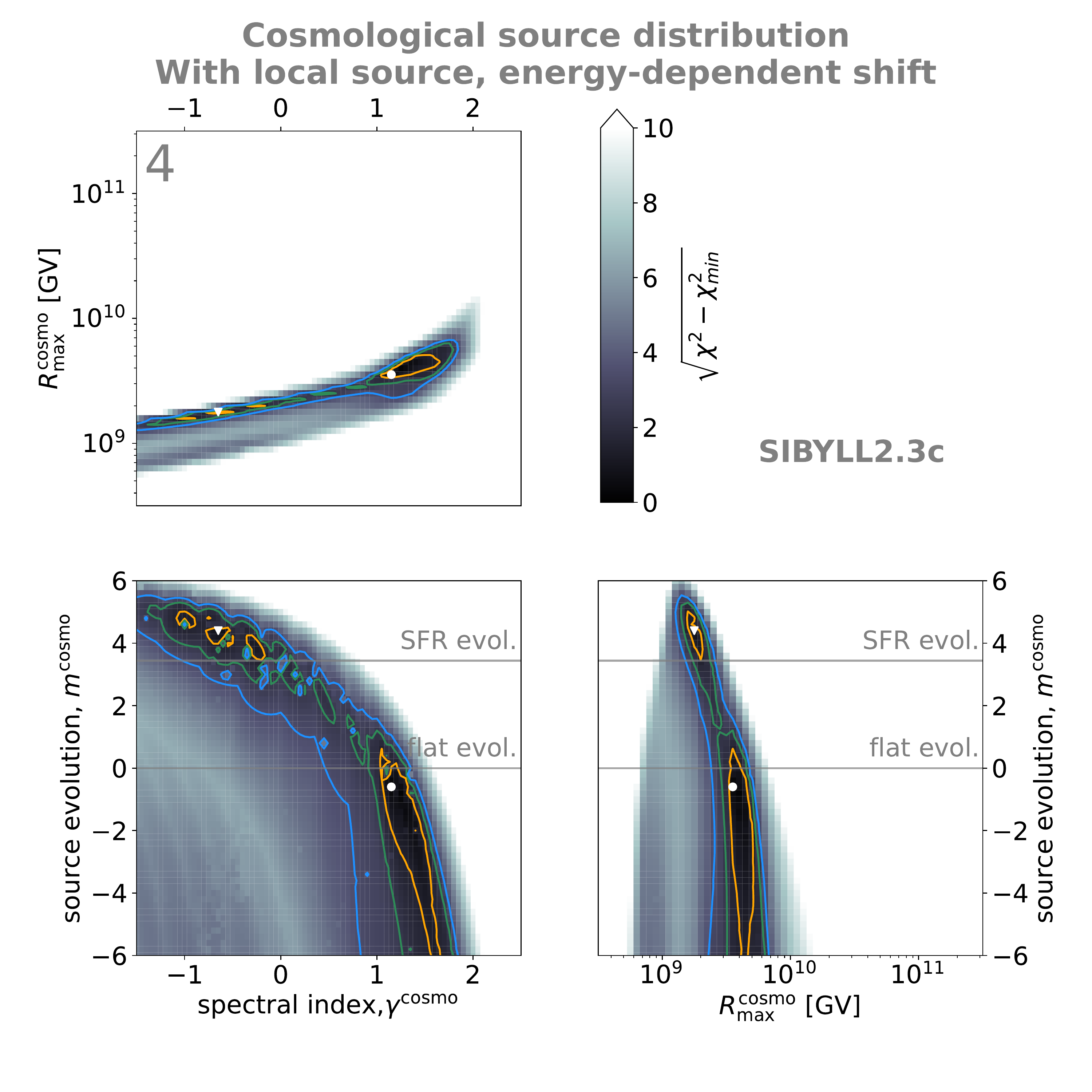}\\
    \end{tabular}
    \caption{Parameter space of a cosmological source distribution based on a joint fit to TA and PAO data. In the upper left panel we assume an energy-independent shift in the energy scales of the experiments with no local source, while in the upper right panel w also consider the existence of a local source in the Northern Hemisphere (observed only by TA). In the bottom left panel we assume an energy-dependent shift and in the bottom right panel we consider both an energy-dependent shift and the presence of a local source in the Northern Hemisphere. The white dots in each plot correspond to the best-fit parameters of each scenario (also listed in \Fig\ref{fig:main_result_predictions}), while a white triangle represents a second minimum. The colored shading corresponds to the $\chi^2$ value compared to the best fit, while the yellow, green and blue contours indicate the 1-, 2- and $3\sigma$~regions, respectively, calculated for two d.o.f. In each panel, the parameter that is not shown is treated as a nuisance parameter and minimized over. \textsc{Sibyll 2.3}c was used as the air shower model.}
    \label{fig:main_result_contours_cosmo}
\end{figure*}

\begin{figure*}[htpb!]
    \centering     
    \includegraphics[width=.45\textwidth]{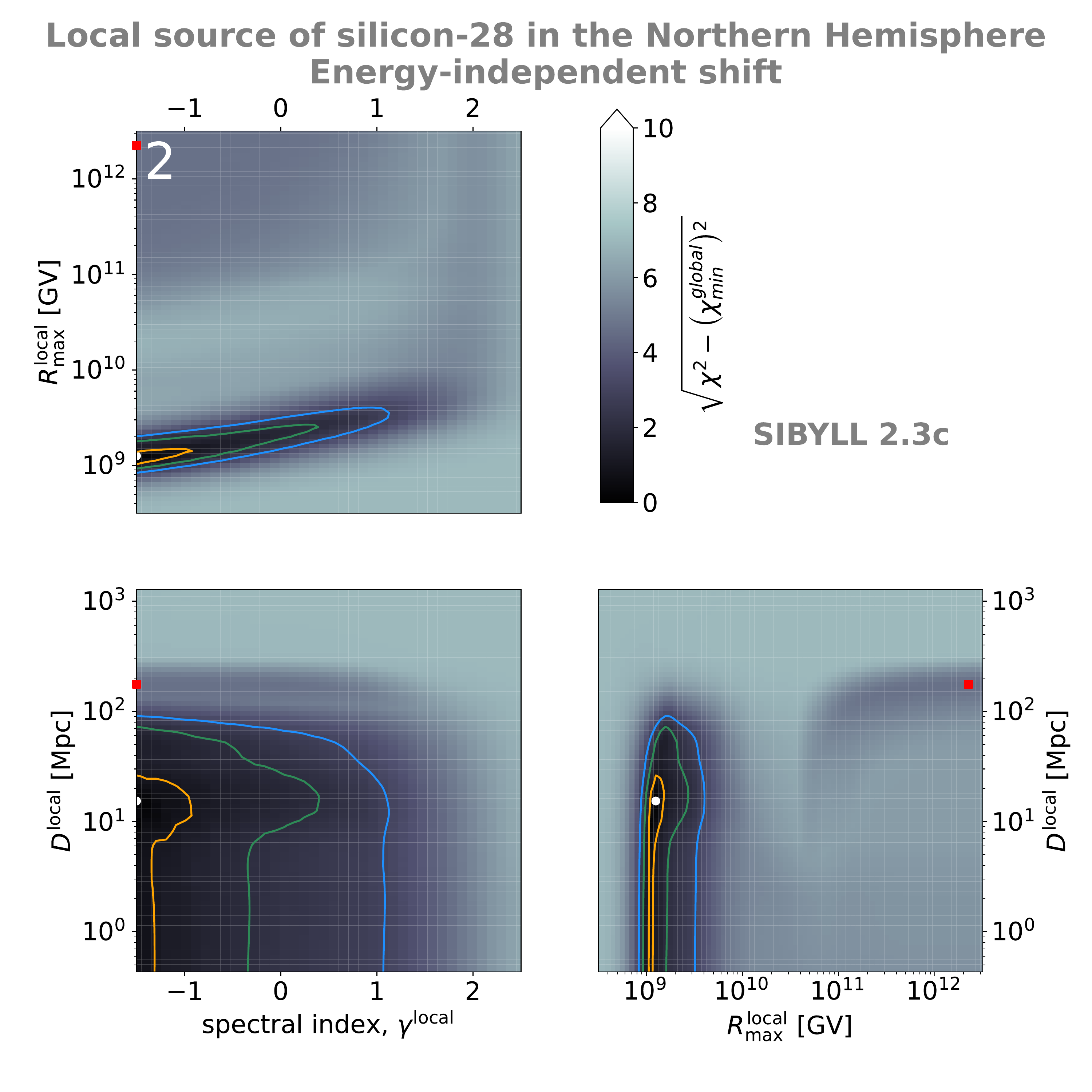} \includegraphics[width=.45\textwidth]{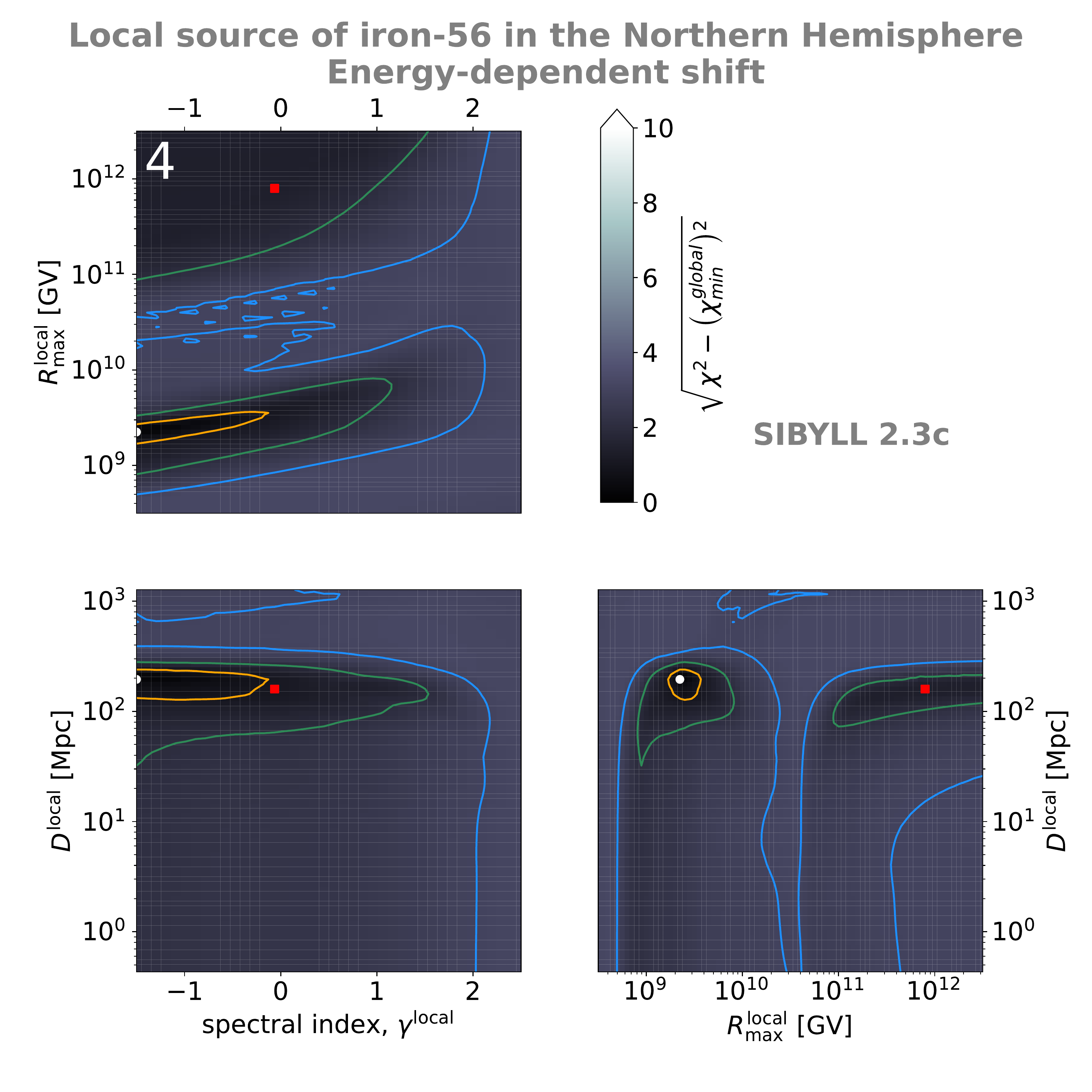}
\caption{
Parameter space of a local source for a joint fit to PAO and TA data when considering a cosmological source distribution and a local source in the Northern Hemisphere. In the left panel, we allow only for energy-independent shifts in the energy scales of the experiments and the local source emits cosmic rays of the silicon-28 mass group, which provides the best fit in that scenario. In the right panel, we allow for energy-dependent systematic shifts and the local source emits cosmic rays of the iron-56 mass group. The best-fit parameters are shown as white dots; a second minimum, discussed in \App\ref{app:exotic}, is marked with red squares. The meaning of the contour colors is as in \Fig\ref{fig:main_result_contours_cosmo}.}
\label{fig:main_result_contours_local_source}
\end{figure*}

\begin{deluxetable*}{c|l|r|r||r|r}
    \centering
    \caption{Best-fit parameters corresponding to the results of the joint fit to PAO and TA data, using \textsc{Sibyll 2.3}c as the air shower model. The 1$\sigma$ uncertainty region is given for 1 d.o.f.}
    \label{tab:main_result_parameters}
    \tablehead{
        & 
        & \multicolumn{2}{c||}{ energy-independent shift}
        & \multicolumn{2}{c}{ energy-dependent shift}
         \\
        &
        &    no local source (1)
        &    with local source (2)
        &    no local source  (3)
        &    with local source (4)
    }

    \startdata
    \multirow{9}{*}{\rotatebox[origin=c]{90}{Cosmological source distrib.}}
    &\gammacosmo
        & $1.60_{-0.05}^{+0.05}$
        & $1.60_{-0.05}^{+0.05}$
        & $1.35_{-0.10}^{+0.10}$
        & $1.15_{-0.05}^{+0.30}$
    \\
    &\Rmaxcosmo~(GV)
        &$5.0_{-0.3}^{+0.3}\times10^{9}$
        &$5.0_{-0.3}^{+0.3}\times10^{9}$
        &$4.5_{-0.5}^{+0.6}\times10^{9}$
        &$3.55_{-0.3}^{+0.9}\times10^{9}$
    \\
    &\mcosmo
        &$<-5.4$
        &$<-5.8$
        &$-1.2_{-1.6}^{+0.6}$
        &$-0.6_{-2.6}^{+0.8}$
    \\
    &$I^9_A (\%)$
        & $\,$
        & $\,$
        & $\,$
        & $\,$
    \\
    &\multicolumn{1}{r|}{H}
        & $0.0_{-0.0}^{+17.1}$
        & $0.0_{-0.0}^{+12.6}$
        & $0.0_{-0.0}^{+99.8}$
        & $0.0_{-0.0}^{+7.5}$
    \\
    &\multicolumn{1}{r|}{He}
        & $6.5_{-4.0}^{+9.2}$
        & $0.0_{-0.0}^{+99.6}$
        & $22.4_{-5.8}^{+7.0}$
        & $44.7_{-3.1}^{+3.2}$
    \\
    &\multicolumn{1}{r|}{N}
        & $75.9_{-1.2}^{+1.2}$
        & $85.1_{-0.4}^{+0.4}$
        & $38.1_{-3.3}^{+3.4}$
        & $29.1_{-2.5}^{+2.6}$
    \\
    &\multicolumn{1}{r|}{Si}
        & $16.8_{-2.3}^{+2.5}$
        & $14.1_{-2.4}^{+2.8}$
        & $38.4_{-1.7}^{+1.7}$
        & $24.8_{-1.3}^{+1.4}$
    \\
    &\multicolumn{1}{r|}{Fe}
        & $0.8_{-0.5}^{+1.1}$
        & $0.8_{-0.5}^{+1.5}$
        & $1.2_{-0.5}^{+1.0}$
        & $1.4_{-0.4}^{+0.5}$
    \\
    \hline
    \parbox[t]{2mm}{\multirow{4}{*}{\rotatebox[origin=c]{90}{Local source}}}
    & isotope 
        & $\,$
        & silicon-28
        & $\,$
        & iron-56
    \\
    &\gammalocal
        & $\,$
        &$<-1.0$
        & $\,$
        &$<-0.2$
    \\
    &\Rmaxlocal~(GV)
        & $\,$
        & $1.3_{-0.1}^{+0.2}\times 10^{9}$
        & $\,$
        & $2.3_{-0.5}^{+1.3}\times 10^{9}$
    \\
    &\lumlocal~(erg $s^{-1}$)
        & $\,$
        & $<3.7\times 10^{42}$
        & $\,$
        & $2.2_{-1.5}^{+1.8}\times 10^{44}$
    \\
    &\Dlocal~(Mpc)
        & $\,$
        & $<25.6$
        & $\,$
        & $194.9_{-64.9}^{+43.6}$
    \\
    \hline
    \multirow{4}{*}{\rotatebox[origin=c]{90}{Systematics}}
    &\deltaEPAO (\%)
        &$-11.92_{-0.05}^{+2.95}$
        &$-13.04_{-0.02}^{+0.01}$
        & $0.40_{-0.08}^{+0.07}$
        & $0.35_{-0.06}^{+0.12}$
    \\
    &\deltaETA (\%)
        &$-21.00_{-0.00}^{+2.58}$
        &$-21.00_{-0.00}^{+0.02}$
        &$-7.65_{-0.02}^{+0.05}$
        &$-7.65_{-0.02}^{+0.02}$
    \\
     &\deltaMeanXmaxPAO (\%)
        &$-58_{-10}^{+15}$
        &$-58_{-1}^{+14}$
        &$-100_{-0}^{+13}$
        &$-100_{-0}^{+21}$
    \\
    &\deltaMeanXmaxTA (\%)
        &$3_{-5}^{+7}$
        &$4_{-1}^{+7}$
        &$1_{-3}^{+7}$
        &$1_{-5}^{+10}$
    \\
    &\deltaSigmaXmaxPAO (\%)
        &$100_{-17}^{+0}$
        &$100_{-1}^{+0}$
        &$55_{-18}^{+22}$
        &$35_{-15}^{+46}$
    \\
    &\deltaSigmaXmaxTA (\%)
        & $-33_{-7}^{+4}$
        & $-72_{-1}^{+6}$
        &$-58_{-4}^{+6}$
        &$-61_{-5}^{+12}$
    \\
    \hline
    &$\chi^2$/d.o.f.
        &110.6/54
        &67.8/50
        &83.6/54
        &76.0/50
    \\
    \hline
    & $\Delta\,\mathrm{AIC}_\mathrm{c}$
        & $\,$
        & 28.9
        & 27.0
        & 20.7
    \\
    &{\makecell{Favored vis-a-vis\\null hypothesis (1)}}
        & $\,$
        & 5.0$\sigma$
        & 4.8$\sigma$
        & 4.2$\sigma$
    \\
    \hline
     & \multicolumn{1}{r|}{$\left(\chi_{\mathrm{spectrum}}^\PAO \right)^2$}
        &18.8
        &15.7
        &14.2
        &16.1
    \\
    & \multicolumn{1}{r|}{$\left(\chi_{\mathrm{spectrum}} ^\TA \right)^2$}
        &55.2
        &13.4
        &24.5
        &18.0
\\
    \hline
    \enddata
\end{deluxetable*}

The composition observables are shown in the bottom plots of each panel in \Fig\ref{fig:main_result_predictions}. Regarding the cosmological source distribution, as reported in \Tab\ref{tab:main_result_parameters}, for the energy-independent shift the best fit is dominated by nitrogen at the 75-85\% level, followed by silicon.The proton component is absent for the best-fit scenario and the protons in the propagated spectrum at Earth are therefore only secondaries. This is consistent with previous fits to PAO data only by~\citet{Heinze:2019} and~\citet{Aab:2017}. That means that the addition of the TA data and constraints on the predicted spectrum at lower energies do not considerably change the predicted composition emitted by the sources. On the other hand, for the energy-dependent shift, helium, nitrogen, and silicon are present at similar levels due to the effect on lower energies (10-30 EeV), where nitrogen dominates. However, it should be noted that in some cases the content of protons and helium emitted directly by the sources can be difficult to constrain because their maximum energy lies below the minimum energy threshold of our fit. This is due to the assumption of rigidity-dependent maximum energies for the emitted cosmic rays, as described in \Sec\ref{sec:methods}.

Regarding the local source, we can see that its contribution in both cases does not change considerably the expected values of $\langle X_\mathrm{max} \rangle$ and $\sigma(X_\mathrm{max})$ above our fitting threshold (compare dashed black curves and solid brown curves in the right bottom panels of \Fig\ref{fig:main_result_predictions}).  For energy-independent shifts we found that silicon, out of the five isotopes we tested, provides the best overall fit quality for \textsc{Sibyll}, nitrogen for \textsc{Epos-LHC}, and iron for \textsc{QGSJET}. However, scenarios where the local source emits other isotopes are also viable. As discussed later in this section and in more detail in \App\ref{app:other_isotopes}, heavier isotopes, such as iron, lead to results that fit the data only marginally worse than silicon, and require that the local source lies farther from Earth. For energy-dependent shifts, only iron provides the best fit for all air shower models, and other elements are not allowed in the 1$\sigma$ region. In both cases, as shown in \App\ref{app:other_isotopes}, considerably lighter isotopes, such as protons and helium, provide comparatively poor fits to data, because the predicted composition at Earth is too light compared to TA data.

Assuming an energy-independent energy shift we find that the existence of a local source in the Northern Hemisphere is preferred over the model with only a cosmological source distribution at the 5.0$\sigma$ level. In the case of only a cosmological source with an energy-dependent shift, both models explain the data equally good at the 4.8$\sigma$ level. When a local source is added together with an energy-dependent shift, this is preferred at a lower level of 4.2$\sigma$ because the additional four parameters do not significantly improve the $\chi^2$ of the joint fit. The other air shower models show that the addition of a local source with an energy-independent shift has a higher significance than the cases with an energy-dependent shift, as shown in \App\ref{app:other-air-shower-models}. Moreover, for the \textsc{QGSJET} air shower model, using the energy-dependent shift leads to a worse fit than the null hypothesis, since the joint fit fails to fit the PAO data.

In \Fig\ref{fig:main_result_contours_cosmo} and \ref{fig:main_result_contours_local_source} we show the values of $\Delta\chi=\sqrt{\chi^2-\chi^2_\mathrm{min}}$ for a region of the parameter space, where $\chi^2_\mathrm{min}$ is the best-fit chi-squared value. In \Fig\ref{fig:main_result_contours_cosmo} we can see the parameter space of the cosmological source distribution for all scenarios. The white dots represent the best-fit parameters, which correspond to the results of \Fig\ref{fig:main_result_predictions}. In the case of an energy-dependent shift with a local source, there is a second minimum, marked as a white triangle.  We show the results for this case in \App\ref{app:second-minimum}. The yellow, green, and blue contours represent the regions 1-, 2-, and 3$\sigma$ away from the best fit. These parameter spaces are consistent with the results in \citet{Heinze:2019} within 3$\sigma$. 

Notably, the 1$\sigma$ region predicts a negative source evolution and a soft spectral index for the cosmological source distribution for all four scenarios.  In contrast, the 1$\sigma$ results from \citet{Heinze:2019} indicate a positive source evolution and a harder spectral index. The differences between our results and those of \citet{Heinze:2019} are primarily due to our inclusion of upper limits on the spectrum at energies lower than $E_\mathrm{min} = 6\times10^{9} \text{ GeV}$. These constraints mainly affect the flux of secondary protons from disintegration. In the best fit from \citet{Heinze:2019}, disintegration proves more efficient due to a positive source evolution, thus overshooting the observed spectrum for $E < E_\mathrm{min}$. In our case, the best-fit result obtained from the joint fit without the restriction on the spectrum for $E < E_\mathrm{min}$ has a $\chi^2_\mathrm{min}$ value of 109.1, along with parameters similar to \citet{Heinze:2019}. However, including the spectrum restrictions leads to an increase of the $\chi^2_\mathrm{min}$ value for the same parameter set to 116.1. In contrast, the overall best-fit result, which includes the restriction on the spectrum, has a $\chi^2_\mathrm{min}$ value equal to 110.1 and does not overshoot the observed spectrum for $E < E_\mathrm{min}$.

In \Fig\ref{fig:main_result_contours_local_source} we show the parameter space of the local source, assuming silicon-28 and iron-56 emission, respectively, for the cases with an energy-independent and energy-dependent shift. The white dots represent the best-fit result (as on the right-hand side of \Fig\ref{fig:main_result_predictions}). As we can see, in both scenarios the strictest constraint obtained is on the maximum rigidity of the cosmic rays accelerated by the local source, \Rmaxlocal. As detailed further in \App\ref{app:other_isotopes}, the best-fit value of $E^\mathrm{local}_\mathrm{max}=Z_A\,R^\mathrm{}_\mathrm{max}$ does not depend on the isotope (or mix of isotopes) accelerated by the source. Another interesting feature of \Rmaxlocal~is that there are two ranges that can provide a good fit. The lower range, centered around 20~EeV, is the best-fit case shown in \Fig\ref{fig:main_result_predictions} (white dots in \Fig\ref{fig:main_result_contours_local_source}); and for $E_\mathrm{max}>2000$~EeV, the model becomes viable again at the 2$\sigma$ level for the energy-dependent shift. The best fits in this energy range are shown as red squares in \Fig\ref{fig:main_result_contours_local_source}. This second case represents a local source that is an extreme accelerator of cosmic rays up to the ZeV regime. In this scenario, the ZeV cosmic rays from the local source disintegrate efficiently into protons and simultaneously cool down to tens of EeV, leading to a pure-proton component at Earth that explains the TA excess. As we can see in the bottom-right plot of the left and right panels of \Fig\ref{fig:main_result_contours_local_source}, the local source must lie at a distance of at least 100~Mpc in order for this strong cooling to occur. Although for Sibyll this ``exotic'' case is at best $2\sigma$ away from the best-fit case for the energy-dependent shift, when considering \textsc{Epos-LHC} there is in fact a viable solution within the $1\sigma$ region. Moreover, the ``exotic'' case assuming other emitted isotopes, for the energy-dependent scenario, provides a better fit than the standard case. These details are discussed further in \App\ref{app:exotic}.

Regarding the spectral index of the cosmic rays emitted by the local source, we can see that our result can only provide an upper limit for \gammalocal~at $<-1.0$ and $<-0.2$ for the energy-independent and energy-dependent shift cases, respectively. This is because softer spectra would lead to an additional flux at Earth below 20~EeV that would overshoot the observed TA flux. On the other hand, a lower limit of \gammalocal~cannot be obtained because for hard spectra the flux becomes dominated by cosmic rays with energies close to $E_\mathrm{max}$, and therefore the precise shape of the distribution cannot be constrained.

\begin{figure*}[htbp!]
\begin{tabular}{cc}
  \includegraphics[width=.445\textwidth]{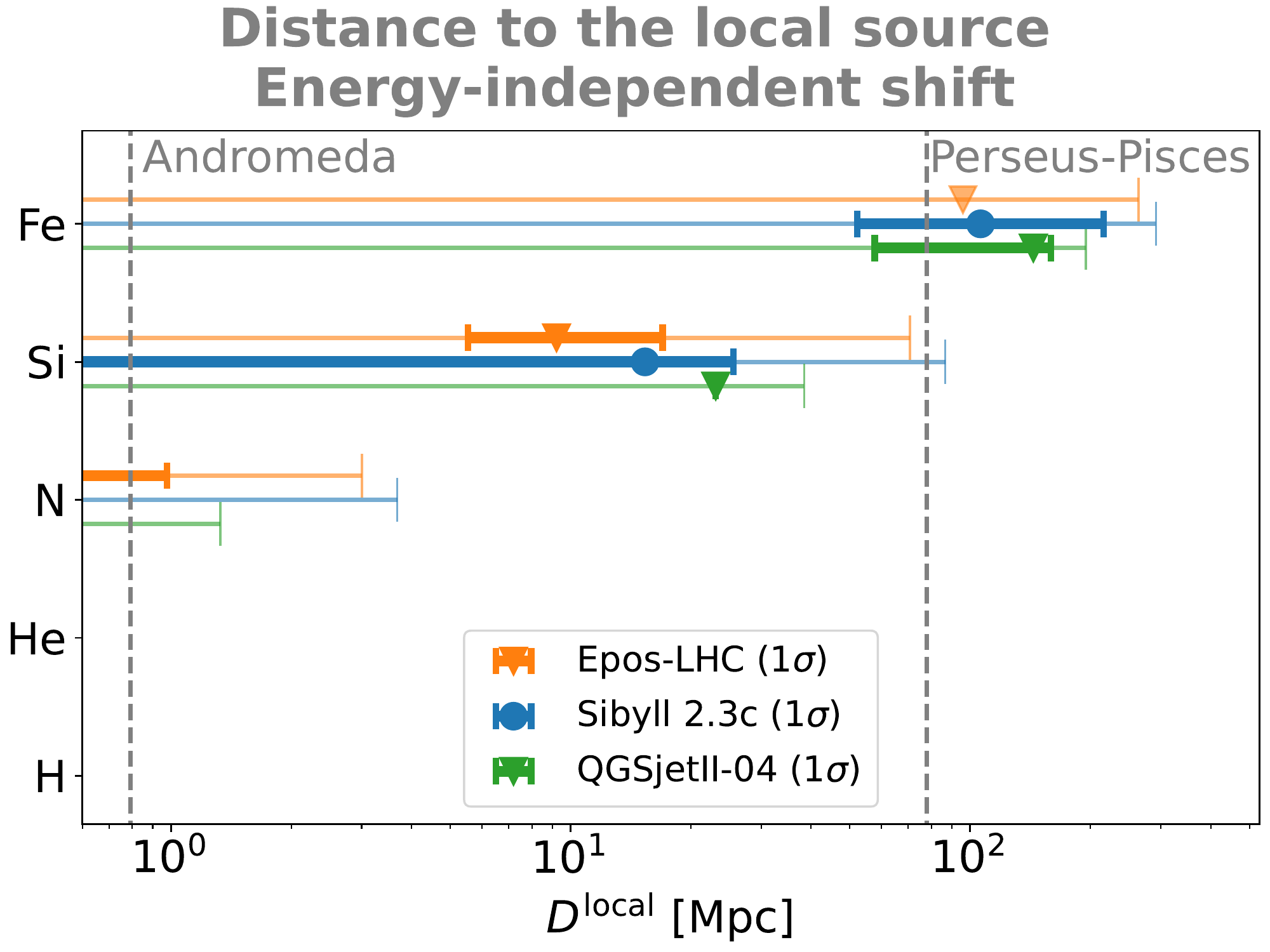}&   \includegraphics[width=.445\textwidth]{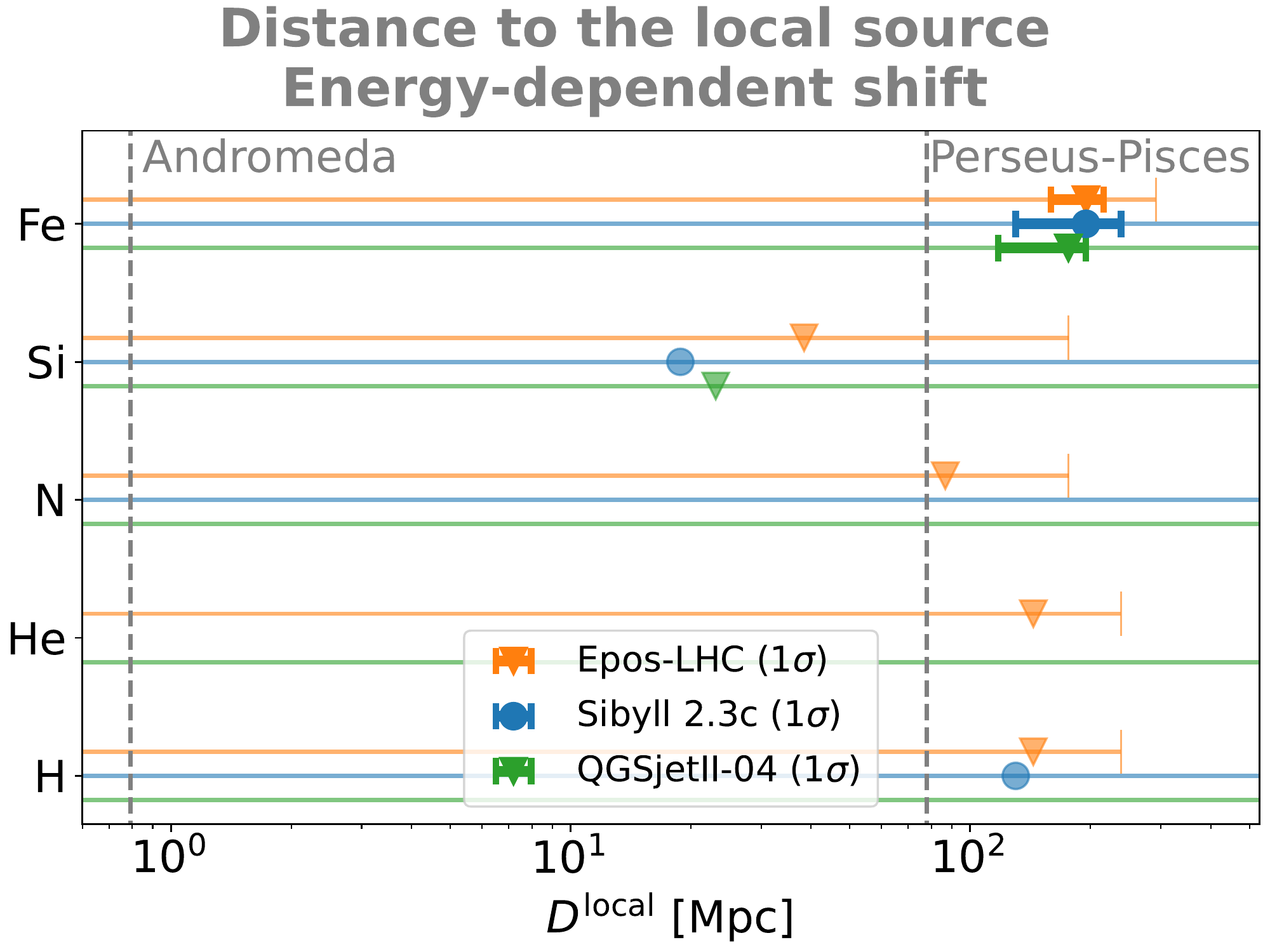} \\
\end{tabular}
\caption{
Best-fit results on the travel distance of the cosmic rays emitted by the local source, depending on the emitted isotope, in the $1\sigma$ region (bold error bars) and $3 \sigma$ (thin error bars), for one degree of freedom. The colors refer to the three different air shower models. The left plot shows results for the energy-independent shift, while the right plot shows results for the energy-dependent shift.}
\label{fig:distance}
\end{figure*}

Finally, we can see that the distance to the local source is also constrained. To better illustrate this we show in the left panel of \Fig\ref{fig:distance} the 1$\sigma$ and 3$\sigma$ uncertainty regions on the distance traveled by the cosmic rays from the local source to Earth, for five different emitted isotopes. The ``exotic'' case is excluded, and all other parameters are kept at their best-fit values. Silicon-28 and iron-56, the isotopes discussed so far, provide the best fits among the five mass groups for the energy-independent and energy-dependent shift cases, respectively. Assuming Sibyll as the air shower model (blue, cf.~also \Tab\ref{tab:main_result_parameters}), the source should lie at any distance below 25.6~Mpc for a result within the $1\sigma$ region for the energy-independent shift.  This constraint on the distance arises from the optimal efficiency of photodisintegration undergone by different nuclei at the highest energies which is necessary to explain the data. For example, the energy loss length of a silicon nucleus with an energy of $\sim$200~EeV is roughly 10~Mpc, which is the same order of magnitude as our optimal local source distance of 14~Mpc. If the local source were to lie much closer to Earth, the emitted silicon nuclei would  undergo less photodisintegration, leading to an observed TA flux with a harder spectrum and heavier composition. However, as we can see in the left-hand plot, in that case we are still able to explain observations within the 1$\sigma$ region of the best fit. On the other hand, if the source were to lie at a distance much larger than the energy loss length, efficient photodisintegration at these energies would be too thorough, producing large amounts of secondary nuclei. These lighter isotopes should then be observed by TA at lower energies (due to their lower mass number), leading to an additional flux that is not supported by the data. For that reason, distances much larger than $\sim10$~Mpc are excluded for the case of silicon. Summarizing the results for the energy-independent shift, for lighter isotopes like nitrogen, the maximum distance to the source is limited to a few~Mpc, while for heavier isotopes like iron, it can be much larger, of the order of 100~Mpc. In the case of the energy-dependent shift, only iron can fit the spectrum data within the 1$\sigma$ region, due to the narrow energy range where differences between the TA and PAO flux occur in this case.

As shown in \Fig\ref{fig:distance}, Andromeda (M31) lies at a distance of $752\pm27$~kpc~\citep{Riess_2012} and within the contour region of the new excess of TA. Our neighboring galaxy is therefore  a potential local UHECR source candidate for intermediate-mass isotopes like nitrogen and silicon. However, note that being a spiral galaxy, this seems to be a bold claim as the question of the acceleration sites of such energetic cosmic rays cannot be easily addressed in this case. On the other hand, a source such as the Perseus-Pisces supercluster (PPS, also known as A 426), at 70~Mpc, would satisfy the distance criterion for isotopes in the iron group. Both these objects are also supported as possible local source candidates by current data from TA, since their position is compatible with the direction of the high-energy excess recently detected~\citep{Kim:2021Aj}. Additionally, other candidate sources may also of course exist. Possible candidates may even lie outside the region of the new excess, due to cosmic-ray deflections by the Galactic magnetic fields (GMFs); see also the discussion below.

In terms of energetics, as detailed in \Tab\ref{tab:main_result_parameters} and \Tab\ref{tab:other_isotopes} in \App\ref{app:other_isotopes}, the cosmic-ray luminosity required for a nitrogen source is at the $10^{39}$~erg/s level, while for an iron source, the required luminosity is higher at the $10^{43}$~erg/s. Both Andromeda and the PPS have higher X-ray luminosity in the 0.1--2.4 keV energy band ($10^{44}$~erg/s and $8\times10^{44}$~erg/s~\citep{Boehringer:2021mix}, respectively), making them potentially feasible candidates within this model from the energetics point of view. In the case of Andromeda, a very low cosmic-ray loading of $10^{-5}$ would be sufficient to explain observations; such a source would be barely detectable as an astrophysical neutrino point source even if the cosmic-ray interactions were efficient~\citep{IceCube:2019cia}.

In general, large-scale structured GMFs will cause a shift in the position in the sky of the UHECR source. On the other hand, the presence of small-scale turbulent GMFs leads to a spreading effect around the source position (see e.g.~\citet{Shaw:2022lqd}). 
This spreading effect needs to be small enough for the local UHECR source in the Northern Hemisphere in order not to affect our declination-dependent interpretation of the spectrum.
The amount of spread around the (potentially shifted) source position can be estimated as (see e.g.~\citet{Lee:1995tm}):
\begin{align}
    \theta_\text{rms} \approx & \frac{2}{\pi} \sqrt{D \lambda_B} \frac{ZeB}{E} \nonumber\\
    \approx & 0.35^\circ Z \left( \frac{D}{10~\text{kpc}} \right)^{1/2} \left( \frac{\lambda_B}{100~\text{pc}} \right)^{1/2} \nonumber\\
    & \times \left( \frac{B}{1~\mu\text{G}} \right) \left( \frac{E}{100~\text{EeV}} \right)^{-1},
    \label{eq:GMFspreading}
\end{align}
with $D$ the distance through the Galaxy where these turbulent magnetic fields are present, $\lambda_B$ the correlation length of the turbulent magnetic fields and $B$ the magnetic-field strength. For silicon-28 with an energy of 30~EeV, and typical order-of-magnitude magnetic-field parameters of $D = 10$~kpc, $\lambda_B = 100$~pc and $B = 1~\mu$G, this gives $\theta_\text{rms} \approx 16^\circ$, which is roughly consistent with the extent of the new excess found by TA. However, note that the extent of the turbulent magnetic fields in our Galaxy, their correlation length and their strength are not well known. If they happen to be significantly stronger or wider spread than the order-of-magnitude estimates given here, the expected UHECR spreading increases significantly. In such a scenario, a lighter composition than silicon might be favorable.     

In addition, we adopt a residual intensity (RI) method (see \citet{Abbasi_2020Anisotropy}) in order to ensure that our model is consistent with the observed anisotropy results, which are shown in the \Tab\ref{tab:result_anisotropy}. The RI is defined in our case as the difference between the number of cosmic rays above $E^{\mathrm{RI}}_{\mathrm{min}}$ from the local source divided by the number of cosmic rays from the cosmological source distribution alone:
\begin{align}
    \mathrm{RI}(E^{\mathrm{RI}}_{\mathrm{min}}) = \frac{N_{\local} (E > E^{\mathrm{RI}}_{\mathrm{min}}) }{N_{\cosmo}(E > E^{\mathrm{RI}}_{\mathrm{min}})}.
    \label{eq:residualintensity}
\end{align}

We calculated the RI values for the hotspot and the new excess using data from \citet{Kim:2021Aj}. The results are shown in the \Tab\ref{tab:result_anisotropy}. Our model predicts an RI above $8.8 \,\mathrm{EeV}$ at a level of 0.01, which is consistent with the results ($\mathrm{RI} <0.1$) for the full sky. For the new excess and the hotspot region, the results for the energy-dependent shift are generally lower, while those for the energy-independent shift are at similar levels. 

We note, however, that the model estimates of the RI are based on our 1D simulations where we know which particles come from the local source and which come from the cosmological population of sources. An experimentally observed RI for the corresponding situation, on the other hand, depends on how much the UHECRs from the local source are deflected in Galactic and extragalactic magnetic fields. The model estimates of the RI should, therefore, only be regarded as rough predictions for observed RIs. For a more detailed analysis it is necessary to account for the deflections of UHECRs, which depend on the strength and directions of the different magnetic fields. Such an analysis goes beyond the scope of this paper.

\begin{deluxetable*}{l|l|r|r|r}[htbp!]
    \centering
    \caption{Comparison of the UHECR anisotropy predicted by the model with the observed values. In the first column we report the minimum energy considered for each data set, in the second column the observed residual intensity (RI) values \citep{Abbasi_2020Anisotropy, Kim:2021Aj}, and in the two right-most columns the values predicted by the best-fit parameters of the model considered in this work. The uncertainties on all the model estimates are of the order of 1\%.} The residual intensity  values for the hotspot and new excess are calculated using data from \citet{Kim:2021Aj}.
    \label{tab:result_anisotropy}
    \tablehead{
        &  Minimum energy (EeV)  
        &  Observed values of RI
        &  \multicolumn{2}{c}{Model estimates of RI}
        \\
        & $\,$
        & $\,$
        & Energy-independent shift (2)
        & Energy-dependent shift (4)
    }
    \startdata
    Full sky 
    &\multicolumn{1}{r|}{8.8}
        & $<0.1$
        & 0.03
        & 0.01
    \\
    \hline
    New excess
        & $\,$
        & $\,$
        & $\,$
    \\
    &
    \multicolumn{1}{r|}{25.1}
        & 0.72 
        & 0.34
        & 0.06
    \\
    &
    \multicolumn{1}{r|}{31.6}
        & 0.87
        & 0.57
        & 0.12
    \\
    &
    \multicolumn{1}{r|}{39.4}
        & 1.09 
        & 1.21
        & 0.31
    \\
    \hline
    Hotspot
        & \multicolumn{1}{r|}{57.0}
        & 1.74
        & 2.3
        & 0.55
    \enddata
\end{deluxetable*}

\section{Summary and conclusions}
\label{sec:conclusion}

We have performed a joint fit to current PAO and TA data accounting for both spectra and composition observables. We have considered the standard rigidity-dependent maximum energy assumption and considered all relevant parameters and their multi-parameter correlations. We used energy-independent and energy-dependent shifts to account for the energy systematics of the two experiments. For each shift we have tested the hypothesis that both experiments observe a cosmological source distribution of equal UHECR sources, while TA observes an additional local UHECR source located in the Northern Hemisphere. We have demonstrated that the presence of the local source is favored at the $5.0\sigma$ level compared to the null-hypothesis scenario where there is no local source, assuming an energy-independent shift. The case of an energy-dependent shift with no local source is favored at $4.8\sigma$ level. Note, however, that the additional presence of a local UHECR source does not improve the quality of the fit, if the energy-dependent shifts are present, due to the four additional parameters of the local source to the joint fit. One may speculate that both hypotheses may work equally well; such systematics, however, require physical motivation.  

Regarding the  cosmological source distribution, the parameters resulting from our joint fit are consistent across all four scenarios. The results predict a negative source evolution, such as in the case of tidal disruption events \citep[TDEs,][]{Biehl_2018}, and spectral indices compatible with Fermi acceleration. Notably, the differences between our results and those previously reported by \citet{Heinze:2019} are primarily due to the inclusion of upper limits on the spectrum at energies lower than $E_\mathrm{min} = 6\times10^{9} \text{ GeV}$. Concerning the composition results, the best fits for the energy-independent shift align with the previous fits from \citet{Heinze:2019}. However, in the case of the energy-dependent shift, helium, nitrogen, and silicon appear at comparable levels.

Our best joint fit, for the energy-independent shift, reveals a local source that emits cosmic rays dominated by the silicon-28 mass group, with a hard spectrum (\gammalocal~$<-1.0$) and a maximum energy of \Emaxlocal=20~EeV. Although the best fit is obtained considering Sibyll~2.3c as the air shower model, good fits are also obtained considering QGSJET-II-04 and Epos-LHC. Besides silicon, other isotopes with masses between nitrogen and iron are also viable within the 3$\sigma$ region relative to the best fit (cf.~\App\ref{app:other_isotopes}). In the silicon scenario, the source must lie within a distance of 14~Mpc, making Andromeda a viable candidate. For heavier elements such as iron, the local source should lie at a distance of the order of 100~Mpc, compatible with an object such as the Perseus-Pisces supercluster.  For the energy-dependent shift, the best fit for all air showers is an iron source at a distance of 190~Mpc. This distance is higher compared to the energy-independent shift results for iron due to the more complex shape of the differences between TA and PAO data at high energy. The use of TA spectrum data from the Northern Hemisphere would affect the distance results. Both source candidates have a photon luminosity higher than the cosmic-ray luminosity required by the model, making them energetically viable.
Furthermore, both lie within the angular uncertainty region of the flux excess recently reported by TA~\citep{Kim:2021Aj}; our model predicts that their contribution should be significant above 30 EeV, which is approximately the energy range of the TA excess.

We conclude that a local UHECR source provides a good description for the long-standing discrepancy of the spectrum data between PAO and TA. While our $5.0\sigma$ significance with respect to the null hypothesis (cosmological source distribution only, standard systematics) is similar to $4.8\sigma$ for a systematical one, an astrophysical explanation is clearly more attractive from our point of view. Claiming a discovery would require a) a better understanding of possible energy-dependent systematics, b) a scrutinizing analysis performed by the experimental collaborations using updated data, and c) an unambiguous association with (possibly observed) anisotropies 4) performing a joint fit for different declination bands. On the modeling side, have we restricted ourselves to a single mass group from the local source due to the computational effort, while a more complex model involving a mix of isotopes may eventually provide a better joint fit, and could constrain the properties of the local source further. However, the higher number of parameters of such a model will require higher statistics from the Northern Hemisphere, which can only be made possible by future experiments such as the planned TAx4 experiment~\citep{TelescopeArray:2021jim}.

\section*{Acknowledgments}

The authors would like to thank Teresa Bister, Domenik Ehlert, Ralph Engel, Simone Garrappa, Tim L. Holch, Cecilia Lunardini, Ioana Maris, Andrew M. Taylor, and Michael Unger for helpful comments and discussions. PP was supported by the International Helmholtz-Weizmann Research School for Multimessenger Astronomy, largely funded through the Initiative and Networking Fund of the Helmholtz Association. XR was supported by the German Science Foundation DFG, via the Collaborative Research Center SFB1491 ``Cosmic Interacting Matters - From Source To Signal''.

% \bibliography{bibliography.bib}
% \bibliographystyle{aasjournal}
    
\clearpage

\appendix
\section{Results of the joint fit to PAO and TA for the \textsc{Epos-LHC} and \textsc{QGSJET-II-04} air shower models}
\label{app:other-air-shower-models}

In the main text we presented the results of our joint fit to PAO and TA data using the \textsc{Sibyll 2.3}c air shower model. In this section we extend our analysis to the \textsc{Epos-LHC} and \textsc{QGSJET-II-04} models, for which the best-fit parameters are listed in \Tab\ref{tab:result_parameters_epos} and \ref{tab:result_parameters_qgsjet}, respectively.  As noted by \citet{Heinze:2019}, the choice of air shower model can strongly affect the best-fit parameters of the cosmological source distribution, making it essential to examine all available models. The best-fit parameters of the cosmological source distribution are broadly consistent with those presented in \citet{Heinze:2019}; however, we note a discrepancy concerning the composition. Specifically, our results for \textsc{Epos-LHC} and \textsc{QGSJET-II-04} suggest that the composition is predominantly helium (50-70\%) followed by nitrogen, while \citet{Heinze:2019} predict a higher proportion of nitrogen. This discrepancy arises because we include lower-energy spectra as upper limits in our model, which is necessary to prevent a secondary peak of protons from nitrogen that exceeds the data at energies below $6\times10^{9} \text{ GeV}$.

The energy-independent systematic scenario with a local source is preferred over the null hypothesis at the 5.0$\sigma$ level for both \textsc{Epos-LHC} and \textsc{Sibyll 2.3}c. In the case of \textsc{Epos-LHC}, the joint fit has a value of $91.2/50=1.8$ per degree of freedom (d.o.f.), which is similar to the PAO-only fit by~\citet{Heinze:2019}, who obtained a lower $\chi^2$/d.o.f. of 2.2. The value is lower in our case, since we additionally included the spectral data below $E_\{\mathrm{min}$. However, the energy-dependent shift scenarios are less favored due to the upper limit for a nitrogen composition. In the case of \textsc{QGSJET-II-04}, the scenario with energy-independent systematics with a local source is preferred at the level of 2.9$\sigma$. The joint fit has a $\chi^2$ value of $268.5/50=5.4$ per d.o.f., which is better than the PAO-only fit by~\citet{Heinze:2019}, who obtained a lower $\chi^2$/d.o.f. of $12.3$, since the joint fit includes TA data and the spectrum data below $E_{\mathrm{min}}$  Both energy-dependent shift scenarios provide a poor fit, as the joint fit fails to fit the PAO data. Overall, for both air shower models, the scenario of a local source with energy-independent shift provides a better fit to TA and PAO data.

\begin{deluxetable}{c|l|r|r||r|r}
    \centering
    \caption{Best-fit parameters corresponding to the results of the joint fit to PAO and TA data using \textsc{Epos-LHC} as the air shower model. The 1$\sigma$ uncertainty region is given for 1 d.o.f.}
    \label{tab:result_parameters_epos}
    \tablehead{
        & 
        & \multicolumn{2}{c||}{Energy-independent shift}
        & \multicolumn{2}{c}{Energy-dependent shift}
         \\
        &
        &    No local source 
        &    With local source 
        &    No local source
        &    With local source
    }

    \startdata
    \multirow{9}{*}{\rotatebox[origin=c]{90}{Cosmological source distrib.}}
    &\gammacosmo
        & $0.15_{-0.05}^{+0.05}$
        & $0.20_{-0.05}^{+0.30}$
        & $0.45_{-0.05}^{+0.05}$
        & $0.40_{-0.05}^{+0.05}$
    \\
    &\Rmaxcosmo~(GV)
        & $2.51_{-0.3}^{+0.3}\times10^{9}$
        & $2.51_{-0.3}^{+0.3}\times10^{9}$
        & $2.82_{-0.3}^{+0.3}\times10^{9}$
        & $2.82_{-0.3}^{+0.3}\times10^{9}$
    \\
    &\mcosmo
        &$<-5.8$
        &$<-5.6$
        &$<-5.8$
        &$<-6.0$
    \\
    &$I^9_A (\%)$
        & $\,$
        & $\,$
        & $\,$
        & $\,$
    \\
    &\multicolumn{1}{r|}{H}
        & $0.0_{-0.0}^{+100.0}$
        & $0.0_{-0.0}^{+47.6}$
        & $0.0_{-0.0}^{+99.9}$
        & $0.0_{-0.0}^{+2.3}$
    \\
    &\multicolumn{1}{r|}{He}
        & $67.7_{-0.8}^{+0.8}$
        & $67.4_{-0.8}^{+0.8}$
        & $63.3_{-1.1}^{+1.1}$
        & $62.6_{-1.1}^{+1.1}$
    \\
    &\multicolumn{1}{r|}{N}
        & $29.1_{-0.9}^{+1.0}$
        & $29.4_{-1.0}^{+1.0}$
        & $31.1_{-1.2}^{+1.2}$
        & $31.9_{-1.2}^{+1.2}$
    \\
    &\multicolumn{1}{r|}{Si}
        & $3.1_{-0.5}^{+0.7}$
        & $3.1_{-0.6}^{+0.7}$
        & $5.0_{-0.7}^{+0.8}$
        & $5.0_{-0.7}^{+0.8}$
    \\
    &\multicolumn{1}{r|}{Fe}
        & $0.2_{-0.1}^{+0.2}$
        & $0.1_{-0.1}^{+0.2}$
        & $0.6_{-0.1}^{+0.2}$
        & $0.5_{-0.1}^{+0.2}$
    \\
    \hline
    \parbox[t]{2mm}{\multirow{4}{*}{\rotatebox[origin=c]{90}{Local source}}}
    & isotope 
        & $\,$
        & nitrogen-14
        & $\,$
        & iron-56
    \\
    &\gammalocal
        & $\,$
        &$<-1.1$
        & $\,$
        &$<-0.4$
    \\
    &\Rmaxlocal~(GV)
        & $\,$
        & $2.5_{-0.3}^{+0.3}\times 10^{9}$
        & $\,$
        & $2.2_{-0.5}^{+0.9}\times 10^{9}$
    \\
    &\lumlocal~(erg $s^{-1}$)
        & $\,$
        & $<6.6\times 10^{39}$
        & $\,$
        & $2.3_{-1.2}^{+0.9}\times 10^{44}$
    \\
    &\Dlocal~(Mpc)
        & $\,$
        & $<1.0$
        & $\,$
        & $194.9_{-35.7}^{+20.7}$
    \\
    \hline
    \multirow{4}{*}{\rotatebox[origin=c]{90}{Systematics}}
    &\deltaEPAO (\%)
        & $-7.18_{-0.01}^{+0.20}$
        & $-7.08_{-0.03}^{+4.40}$
        & $0.32_{-0.07}^{+0.02}$
        & $0.30_{-0.01}^{+0.10}$
    \\
    &\deltaETA (\%)
        & $-16.85_{-0.01}^{+0.19}$
        & $-15.91_{-0.06}^{+4.16}$
        & $-7.65_{-0.02}^{+0.02}$
        & $-7.65_{-0.02}^{+0.02}$
    \\
     &\deltaMeanXmaxPAO (\%)
        & $-100_{-0}^{+1}$
        & $-100_{-0}^{+1}$
        & $-100_{-0}^{+1}$
        & $-100_{-0}^{+1}$
    \\
    &\deltaMeanXmaxTA (\%)
        & $-19_{-3}^{+1}$
        & $-13_{-3}^{+3}$
        & $-8_{-1}^{+3}$
        & $-5_{-3}^{+1}$
    \\
    &\deltaSigmaXmaxPAO (\%)
        & $-67_{-8}^{+1}$
        & $-75_{-7}^{+14}$
        & $-67_{-1}^{+8}$
        & $-61_{-8}^{+1}$
    \\
    &\deltaSigmaXmaxTA (\%)
        & $-92_{-2}^{+1}$
        & $-100_{-1}^{+1}$
        & $-95_{-1}^{+2}$
        & $-94_{-2}^{+1}$
    \\
    \hline
    &$\chi^2$/d.o.f.
        &134.0/54
        &91.2/50
        &120.4/54
        &108.6/50
    \\
    \hline
    & $\Delta\,\mathrm{AIC}_\mathrm{c}$
        & $\,$
        & 28.9
        & 13.6
        & 11.5
    \\
     &{\makecell{Favored vis-a-vis\\null hypothesis (1)}}
        & null hypothesis
        & 5.0$\sigma$
        & 3.3$\sigma$
        & 2.9$\sigma$
    \\
    \hline
     & \multicolumn{1}{r|}{$\left(\chi_{\mathrm{spectrum}}^\PAO \right)^2$}
        &21.0
        &19.0
        &24.6
        &24.5
    \\
    & \multicolumn{1}{r|}{$\left(\chi_{\mathrm{spectrum}} ^\TA \right)^2$}
        &53.5
        &13.5
        &28.3
        &18.3
\\
    \hline
    \enddata
\end{deluxetable}

\begin{deluxetable}{c|l|r|r||r|r}
    \centering
    \caption{Best-fit parameters corresponding to the results of the joint fit to PAO and TA data using \textsc{QGSJET-II-04} as the air shower model. The 1$\sigma$ uncertainty region is given for 1 d.o.f.}
    \label{tab:result_parameters_qgsjet}
    \tablehead{
        & 
        & \multicolumn{2}{c||}{ energy-independent shift}
        & \multicolumn{2}{c}{ energy-dependent shift}
         \\
        &
        &    no local source 
        &    with local source
        &    no local source 
        &    with local source
    }

    \startdata
    \multirow{9}{*}{\rotatebox[origin=c]{90}{Cosmological source distrib.}}
    &\gammacosmo
        & $-0.70_{-0.05}^{+0.05}$
        & $-0.70_{-0.05}^{+0.05}$
        & $-0.90_{-0.05}^{+0.05}$
        & $-0.90_{-0.05}^{+0.05}$
    \\
    &\Rmaxcosmo~(GV)
        & $2.24_{-0.3}^{+0.3}\times10^{9}$
        & $2.24_{-0.3}^{+0.3}\times10^{9}$
        & $2.00_{-0.3}^{+0.3}\times10^{9}$
        & $2.00_{-0.3}^{+0.3}\times10^{9}$
    \\
    &\mcosmo
        &$<-5.6$
        &$<-5.6$
        &$<-6.0$
        &$<-6.0$
    \\
    &$I^9_A (\%)$
        & $\,$
        & $\,$
        & $\,$
        & $\,$
    \\
    &\multicolumn{1}{r|}{H}
        & $0.0_{-0.0}^{+20.9}$
        & $0.0_{-0.0}^{+6.4}$
        & $11.2_{-4.2}^{+6.3}$
        & $13.0_{-4.1}^{+5.5}$
    \\
    &\multicolumn{1}{r|}{He}
        & $64.7_{-1.0}^{+1.0}$
        & $63.4_{-1.2}^{+1.2}$
        & $59.5_{-2.2}^{+2.2}$
        & $57.1_{-2.3}^{+2.2}$
    \\
    &\multicolumn{1}{r|}{N}
        & $33.3_{-0.8}^{+0.8}$
        & $35.3_{-1.0}^{+1.0}$
        & $25.9_{-1.1}^{+1.2}$
        & $26.6_{-1.1}^{+1.2}$
    \\
    &\multicolumn{1}{r|}{Si}
        & $1.8_{-0.5}^{+0.7}$
        & $1.1_{-0.6}^{+1.3}$
        & $3.2_{-0.5}^{+0.7}$
        & $3.1_{-0.6}^{+0.7}$
    \\
    &\multicolumn{1}{r|}{Fe}
        & $0.2_{-0.1}^{+0.1}$
        & $0.2_{-0.1}^{+0.2}$
        & $0.2_{-0.1}^{+0.2}$
        & $0.2_{-0.1}^{+0.2}$
    \\
    \hline
    \parbox[t]{2mm}{\multirow{4}{*}{\rotatebox[origin=c]{90}{Local source}}}
    & isotope 
        & $\,$
        & iron-56
        & $\,$
        & iron-56
    \\
    &\gammalocal
        & $\,$
        &$<-0.1$
        & $\,$
        &$<-0.3$
    \\
    &\Rmaxlocal~(GV)
        & $\,$
        & $2.2_{-1.2}^{+0.6}\times 10^{9}$
        & $\,$
        & $2.2_{-0.6}^{+0.9}\times 10^{9}$
    \\
    &\lumlocal~(erg $s^{-1}$)
        & $\,$
        & $2.4_{-2.2}^{+1.1}\times 10^{44}$
        & $\,$
        & $1.6_{-1.2}^{+0.6}\times 10^{44}$
    \\
    &\Dlocal~(Mpc)
        & $\,$
        &$143.9_{-86.2}^{+15.3}$
        & $\,$
        & $176.2_{-58.7}^{+18.7}$
    \\
    \hline
    \multirow{4}{*}{\rotatebox[origin=c]{90}{Systematics}}
    &\deltaEPAO (\%)
        & $13.89_{-0.01}^{+0.11}$
        & $13.99_{-0.03}^{+0.01}$
        & $2.54_{-0.01}^{+0.01}$
        & $2.54_{-0.01}^{+0.01}$
    \\
    &\deltaETA (\%)
        & $2.24_{-0.02}^{+0.08}$
        & $3.03_{-0.02}^{+0.02}$
        & $-6.28_{-0.02}^{+0.02}$
        & $-6.46_{-0.02}^{+0.02}$
    \\
     &\deltaMeanXmaxPAO (\%)
        & $-100_{-0}^{+1}$
        & $-100_{-0}^{+1}$
        & $-100_{-0}^{+1}$
        & $-100_{-0}^{+1}$
    \\
    &\deltaMeanXmaxTA (\%)
        & $-54_{-1}^{+1}$
        & $-41_{-2}^{+1}$
        & $-59_{-1}^{+1}$
        & $-59_{-1}^{+1}$
    \\
    &\deltaSigmaXmaxPAO (\%)
        & $100_{-1}^{+0}$
        & $100_{-1}^{+0}$
        & $100_{-1}^{+0}$
        & $100_{-1}^{+0}$
    \\
    &\deltaSigmaXmaxTA (\%)
        & $-2_{-1}^{+1}$
        & $-21_{-2}^{+1}$
        & $-1_{-1}^{+1}$
        & $-0_{-1}^{+1}$
    \\
    \hline
    &$\chi^2$/d.o.f.
        &293.5/54
        &268.5/50
        &349.0/54
        &315.4/50
    \\
    \hline
    & $\Delta\,\mathrm{AIC}_\mathrm{c}$
        & $\,$
        & 11.1
        & -55.5
        & -38.5
    \\
    &{\makecell{Favored vis-a-vis\\null hypothesis}}
        & null hypothesis
        & 2.9$\sigma$
        & excluded by 7.1$\sigma$
        & excluded by 5.6$\sigma$
    \\
    \hline
     & \multicolumn{1}{r|}{$\left(\chi_{\mathrm{spectrum}}^\PAO \right)^2$}
        &64.4
        &63.8
        &72.5
        &72.1
    \\
    & \multicolumn{1}{r|}{$\left(\chi_{\mathrm{spectrum}} ^\TA \right)^2$}
        &55.2
        &30.1
        &37.2
        &29.0
\\
    \hline
    \enddata
\end{deluxetable}

\clearpage
\section{Results for the second minimum of the joint fit to PAO and TA data}
\label{app:second-minimum}

\begin{figure}[htbp!]
    \centering
     \includegraphics[width=.49\textwidth]{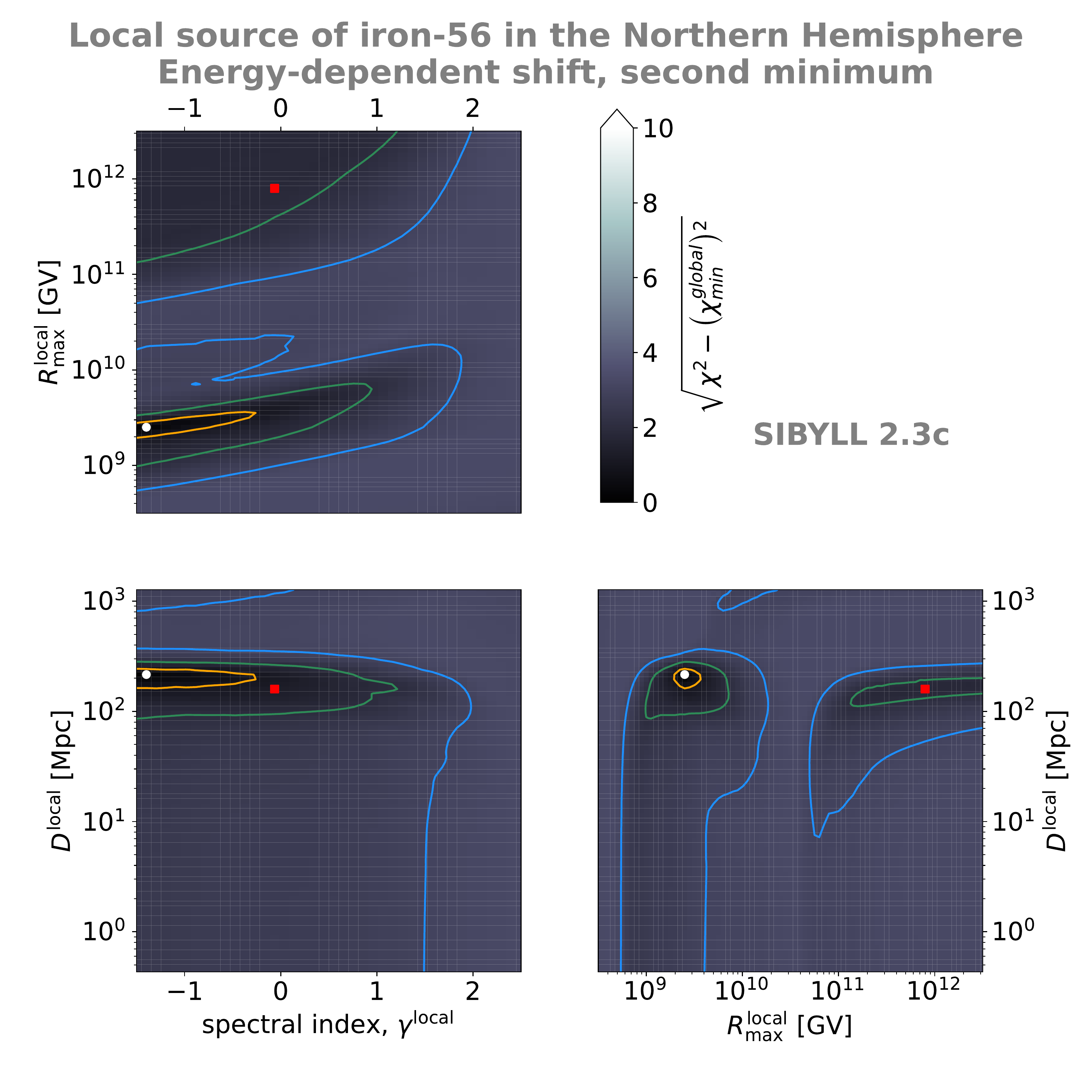}
    \includegraphics[width=.486\textwidth]{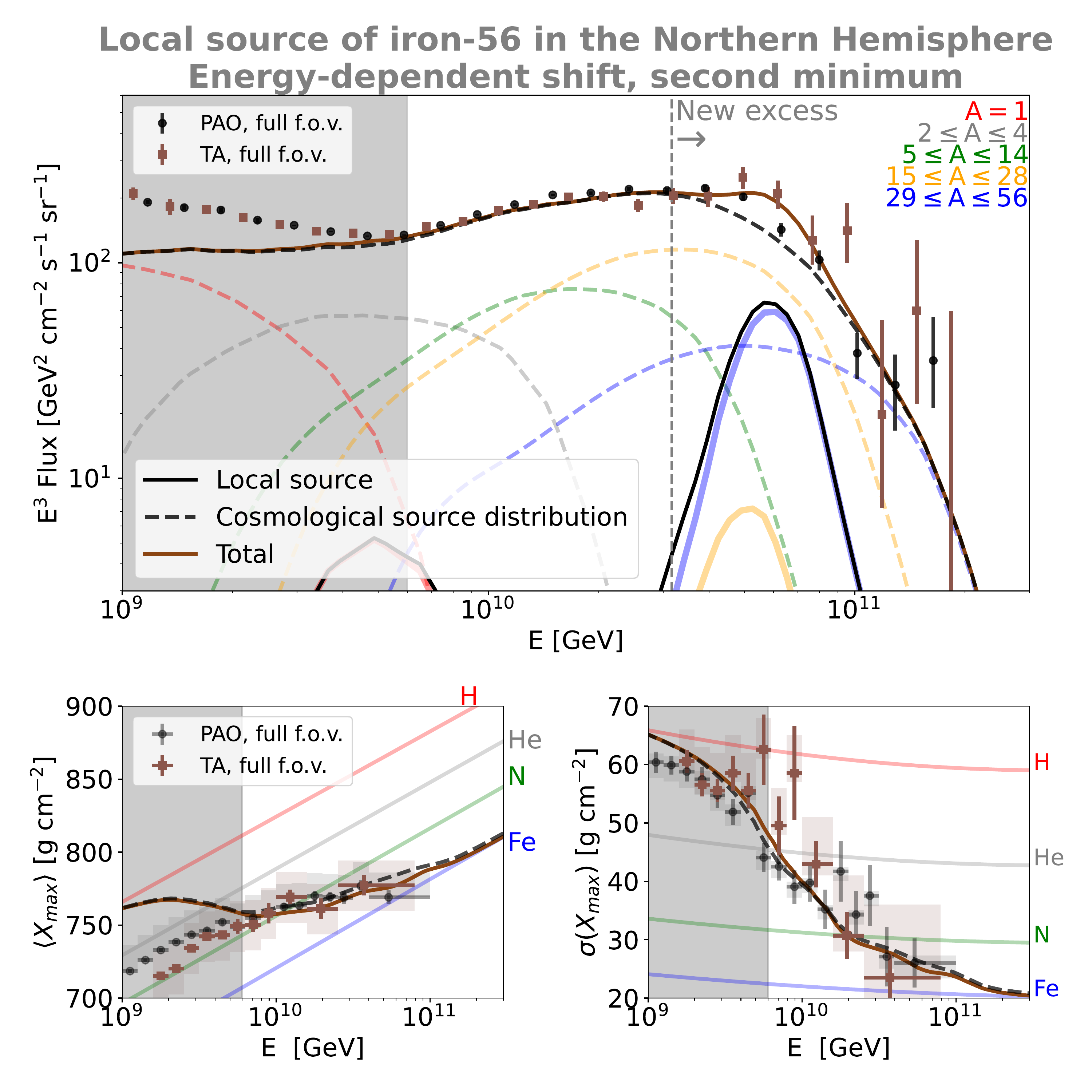}
    \caption{Results representing the second minimum shown as white triangles in \Fig\ref{fig:main_result_contours_cosmo} when considering a cosmological source distribution, a local source in the Northern Hemisphere that emits cosmic rays of the iron-56 mass group, and allowing energy-dependent shifts in the energy scales of the experiments. We use \textsc{Sibyll 2.3}c as the air shower model.  In the left plots we see the parameter space of the local source in this scenario. The best-fit parameters are shown as white dots; a second solution for the local source, discussed in \App\ref{app:exotic}, is marked with red squares. The meaning of the contour colors and gray part is as in \Fig\ref{fig:main_result_contours_cosmo}. In the right plots, spectra (upper panels) and composition observables (lower panels) correspond to the best-fit scenario of the joint fit for the second minimum. The best-fit parameters values are given in \Tab\ref{tab:second_min_result_parameters}.}
    \label{fig:second_minimum_result_predictions}
\end{figure}

As mentioned in the main text, we detect a second minimum in the parameter space of the cosmological source distribution when a local source is present with an energy-dependent shift (see \Fig\ref{fig:main_result_contours_cosmo}). This minimum is similar to the main results of the PAO-only fit by~\citet{Heinze:2019}. The left panel of \Fig\ref{fig:second_minimum_result_predictions} shows the parameter space of a local source for this case, which is very similar to the right plot of \Fig\ref{fig:main_result_contours_local_source}. We detail the best-fit parameters for this case in \Tab\ref{tab:second_min_result_parameters}. The best-fit parameters are also similar for elements other than iron-56. 

As we can see in \Tab\ref{tab:second_min_result_parameters}, the distance to the a local source is slightly higher in this case, but is still compatible with our main result  within the 1$\sigma$ range. All other parameters of the local source are also similar to results for the main minimum. The main difference can be seen in the right panel of \Fig\ref{fig:second_minimum_result_predictions}, where the proton flux at energies 1-10~EV is higher compared to the results for the main minimum. Nonetheless, overall this scenario provides the same level of improvement relative to the null hypothesis as the main result discussed in the main text.

\begin{deluxetable}{l|r||l|r||l|r}
    \centering
    \caption{Best-fit parameters corresponding to the results for the second minimum of the joint fit to PAO and TA. The 1$\sigma$ uncertainty region is given for 1 d.o.f.}
    \label{tab:second_min_result_parameters}
     \tablehead{
        \multicolumn{2}{c||}{Cosmological source distrib.}
        & \multicolumn{2}{c||}{Local source}
        & \multicolumn{2}{c}{Systematics}
    }
    \startdata
    \gammacosmo
        & $-0.65_{-0.05}^{+0.05}$
    &isotope 
        & iron-56
    &\deltaEPAO (\%)
        & $-0.47_{-0.01}^{+0.02}$
    \\
    \Rmaxcosmo~(GV)
        &$1.8_{-0.3}^{+0.3}\times10^{9}$
    &\gammalocal
        &$<-0.4$
    &\deltaETA (\%)
        &$-7.65_{-0.00}^{+0.00}$
    \\
    \mcosmo
        &$4.4_{-0.2}^{+0.2}$
    &\Rmaxlocal~(GV)
        & $2.5_{-0.5}^{+1.1}\times 10^{9}$
    &\deltaMeanXmaxPAO (\%)
        &$-87_{-10}^{+10}$
    \\
    $I^9_A (\%)$
        & $\,$
    &\lumlocal~(erg $s^{-1}$)
        & $3.1_{-1.6}^{+1.0}\times 10^{44}$
    &\deltaMeanXmaxTA (\%)
        &$8_{-4}^{+4}$
    \\
    \multicolumn{1}{r|}{H}
        & $25.9_{-5.6}^{+6.5}$
    &\Dlocal~(Mpc)
        &  $215.6_{-39.4}^{+22.9}$
    &\deltaSigmaXmaxPAO (\%)
        &$-9_{-10}^{+14}$
    \\
    \multicolumn{1}{r|}{He}
        & $0.0_{-0.0}^{+92.5}$
    & $\,$
        & $\,$
    &\deltaSigmaXmaxTA (\%)
        &  $-63_{-3}^{+3}$
    \\
    \multicolumn{1}{r|}{N}
        & $52.8_{-1.6}^{+1.6}$
    & $\,$
        & $\,$
    & $\,$
        & $\,$
    \\
    \multicolumn{1}{r|}{Si}
        &$18.9_{-1.6}^{+1.7}$
    & $\,$
        & $\,$
    & $\,$
        & $\,$
    \\
    \multicolumn{1}{r|}{Fe}
        & $2.3_{-0.4}^{+0.5}$
    & $\,$
        & $\,$
    & $\,$
        & $\,$
    \\
    \hline
    \multicolumn{2}{l}{$\chi^2$/d.o.f.}
    &\multicolumn{4}{c}{76.7/50}
    \\
    \hline
    \multicolumn{2}{l}{$\Delta\,\mathrm{AIC}_\mathrm{c}$}
    &\multicolumn{4}{c}{21.4}
    \\
    \multicolumn{2}{l}{Favored vis-a-vis null hypothesis (1)}
        &\multicolumn{4}{c}{4.1$\sigma$} 
    \\
    \hline
    \multicolumn{2}{l}{$\left(\chi_{\mathrm{spectrum}}^\PAO \right)^2$}
        &\multicolumn{4}{c}{22.9} 
    \\
    \multicolumn{2}{l}{$\left(\chi_{\mathrm{spectrum}} ^\TA \right)^2$}
        &\multicolumn{4}{c}{19.7} 
    \\
   \enddata
\end{deluxetable}

\clearpage
\section{Other isotopes}
\label{app:other_isotopes}

In the main result of \Fig\ref{fig:main_result_predictions} we limited the discussion to the case of silicon-28 and iron-56 for the energy-independent and energy-dependent shift respectively, which provides the best fit. However, the local source may also emit other isotopes, as shown already in \Fig\ref{fig:distance} and discussed in the text thereafter. Here, we summarize the results for those isotopes that do not provide the best fit and are therefore not included in the discussion in the main text. For the sake of simplicity, we discuss only the best-fit  results assuming an energy-independent shift.

\begin{figure*}[htpb!]
    \centering
\begin{tabular}{cc}
   \includegraphics[width=.49\textwidth]{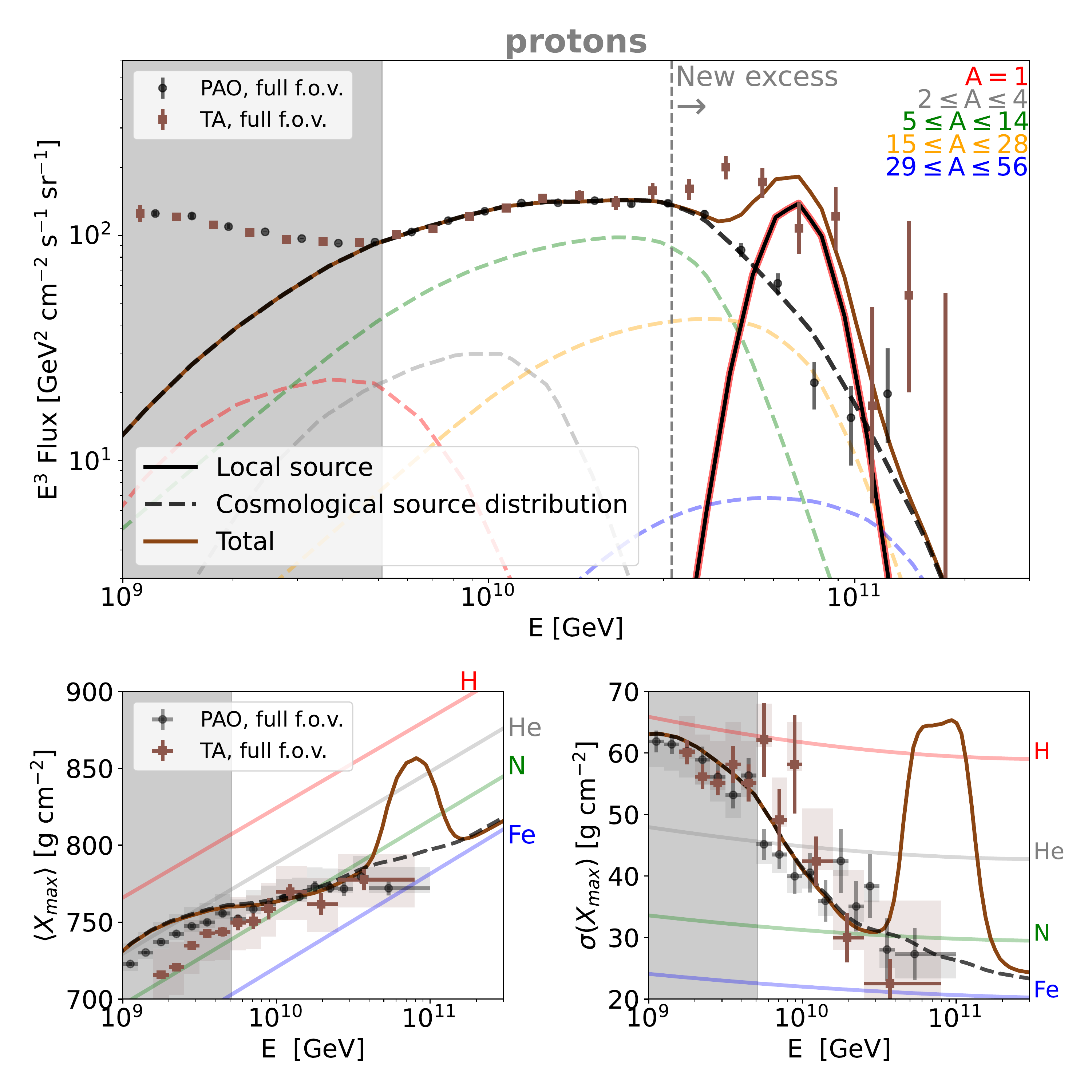} & \includegraphics[width=.486\textwidth]{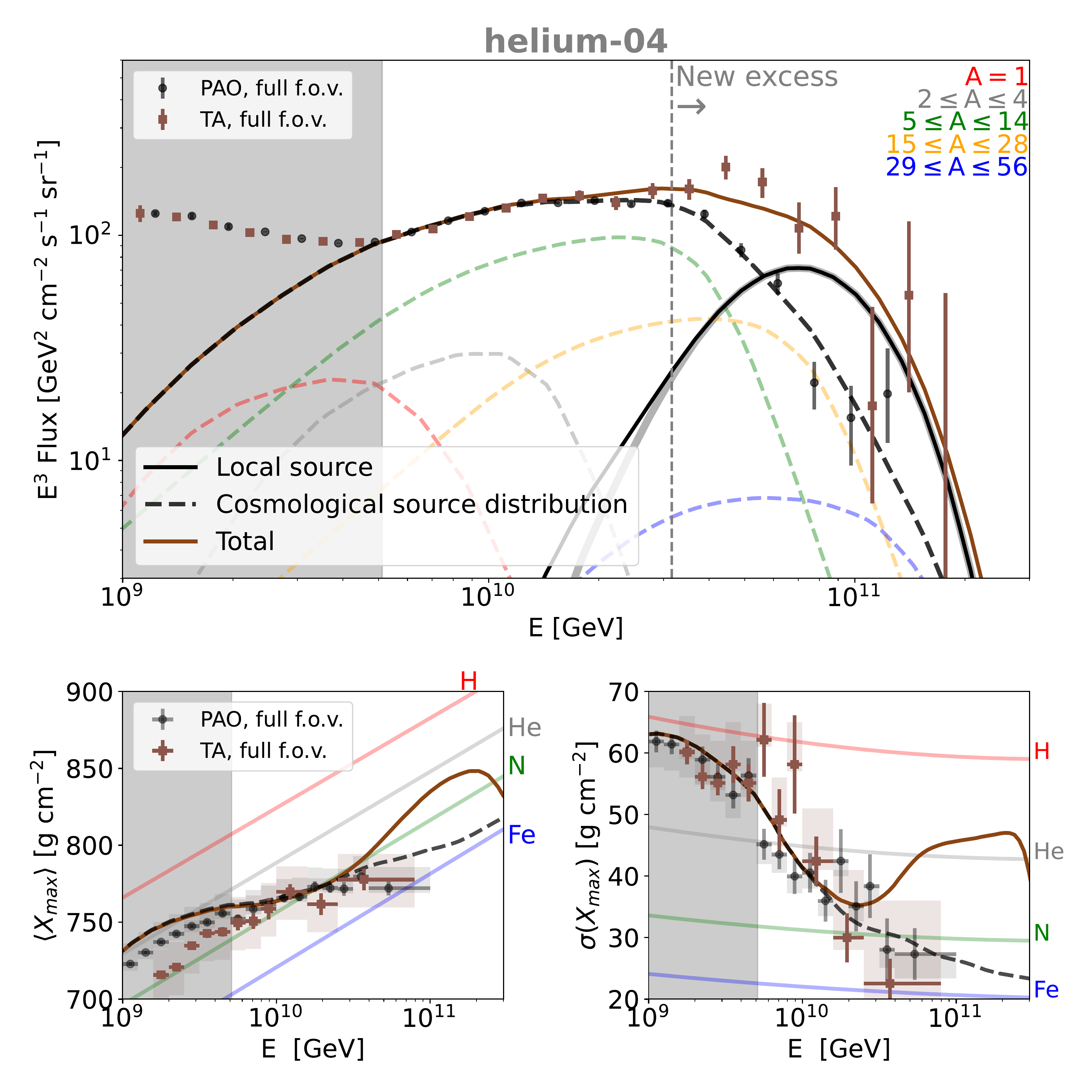} \\
    \includegraphics[width=.49\textwidth]{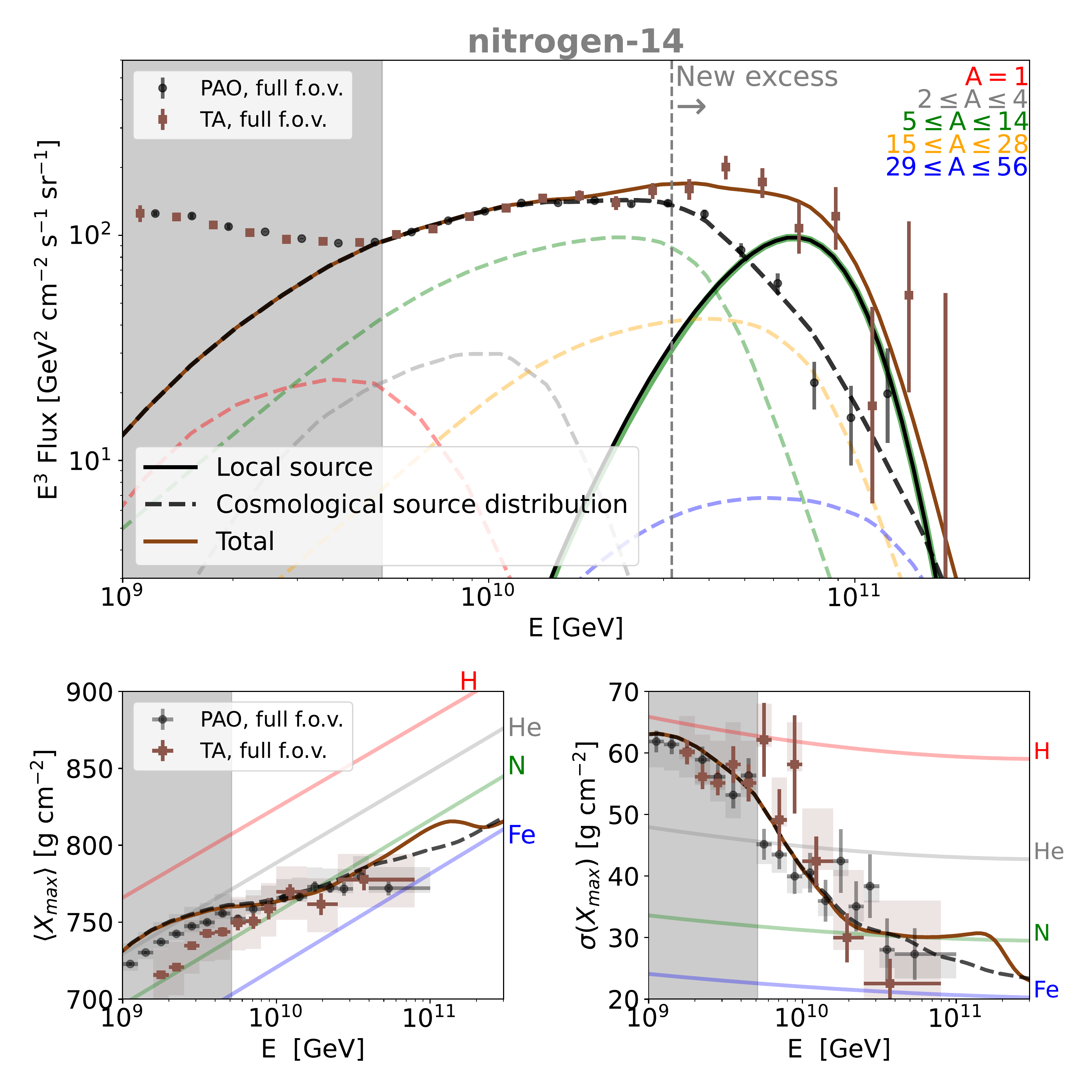}&
    \includegraphics[width=.49\textwidth]{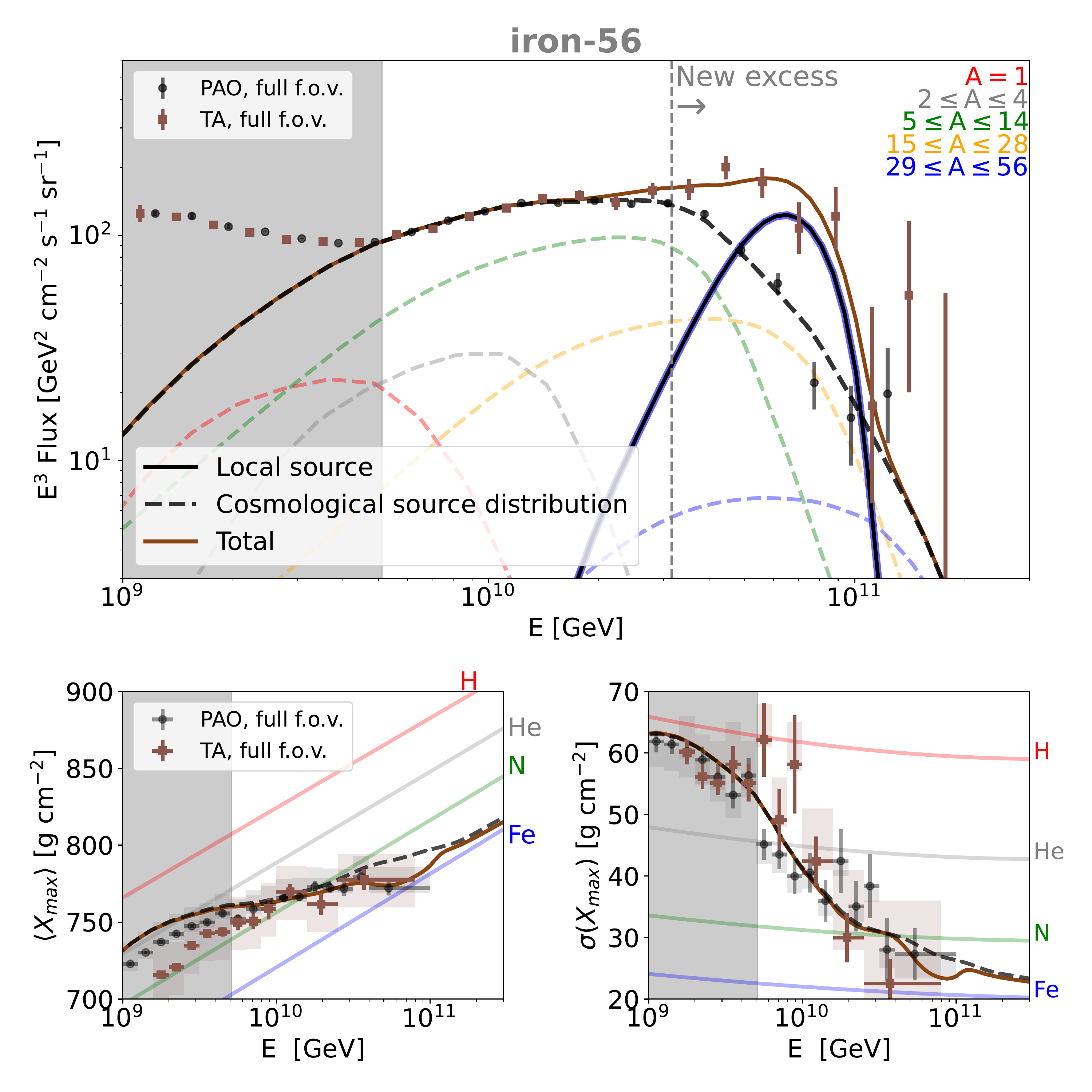}\\
\end{tabular}
\caption{Spectra (upper panels) and composition observables (lower panels) predicted by the models, resulting from a joint fit to TA and PAO data  when  considering a cosmological source distribution and additionally a local source in the Northern Hemisphere, only observable by TA. The plots show the case where this local source emits protons (upper left),  helium-04 (upper right), nitrogen-14 (lower left), and iron-56 (lower right). For the gray shaded energy range, only the spectra data are used as upper limits.}
\label{fig:other_isotopes}
\end{figure*}

In \Fig\ref{fig:other_isotopes} we show the best-fit results for the case where the local source emits protons (upper left), helium-4 (upper right), nitrogen-14 (lower left), and iron-56 (lower right). These results were obtained using Sibyll as the air shower model. Their respective best-fit distances were shown in the main part of this paper as blue points in the left panel of \Fig\ref{fig:distance}. We list the remaining best-fit parameters in \Tab\ref{tab:other_isotopes}, including the best-fit parameters for the other two air shower models that are tested.

As we can see, elements from any mass group up to iron-56 can provide results that are overall compatible with the joint data sets of PAO and TA, provided the emission characteristics and the source distance are adjusted according to \Tab\ref{tab:other_isotopes}. At the same time, some caveats should be noted for the different mass groups, as discussed below.

Emission of either protons or helium-4 leads to an observed composition that is light above 30~EeV, as we can see in the respective composition plots in \Fig\ref{fig:other_isotopes}. However, this does not affect the fit significantly because of the lack of composition data from either TA or PAO at these high energies. The case of a pure proton emission would lead to a result that is qualitatively similar to helium, with even larger values of $\langle X_\mathrm{max}\rangle$ and $\sigma(X_\mathrm{max})$ predicted for TA above 30 EeV.

For intermediate-mass isotopes like nitrogen-14, as well as heavier isotopes with masses up to iron-56, the composition observables exhibit the expected behavior with energy compared to our baseline scenario involving silicon, i.e. the observed composition becomes heavier with energy.

\begin{figure}[htpb!]
\centering
\includegraphics[width=.452\textwidth]{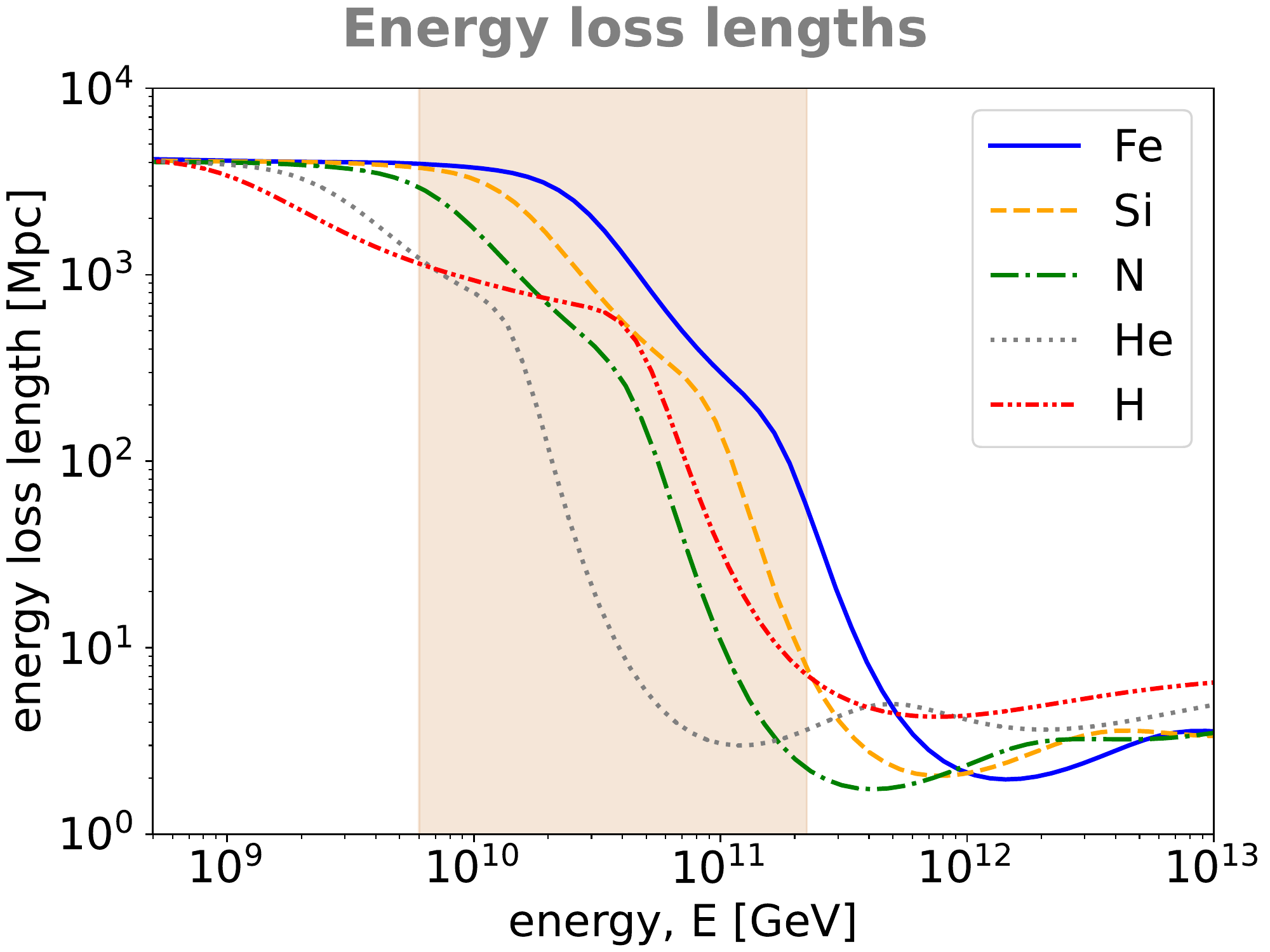} \\
\caption{Energy loss lengths (for redshift $z=0$) of the four tested isotopes as a function of the nucleus energy. This includes adiabatic cooling, photopair production, photopion production and photodisintegration off the CMB and EBL~\citep{Heinze2020Ultra-high-energy}. The orange band shows the energy range relevant for the joint fit.}
\label{fig:energy_loss_lengths}
\end{figure}

In  \Fig\ref{fig:energy_loss_lengths} we show the energy loss length for silicon (yellow curve) and, as a reference, the energy range relevant for our fit (orange band): from 6 EeV, the minimum energy of the joint fit, up to 224~EeV, the highest energy for which TA provides a flux measurement. As we can see, the energy loss length of a silicon nucleus with an energy of $\sim$200~EeV is roughly 10~Mpc, which is the same order of magnitude as our optimal local source distance up to 26~Mpc. If the local source were to lie much closer to Earth, the emitted silicon nuclei would  undergo less photodisintegration, leading to an observed TA flux with a harder spectrum and heavier composition. This is the reason we can provide only an upper limit of the distance of a local source in that case. On the other hand, if the source were to lie at a distance much larger compared to the energy loss length, efficient photodisintegration at these energies would be too thorough, producing large amounts of secondary nuclei. These lighter isotopes should then be observed by TA at lower energies (due to their lower mass number), leading to an additional flux that is not supported by the data. For that reason, distances much larger than $\sim10$ Mpc are excluded for the case of silicon.

By the same token, for a lighter element like nitrogen, the maximum distance to the source becomes limited to about 1~Mpc, as we can see in the \Tab\ref{tab:other_isotopes}, while for heavier isotopes like iron it is much larger, of order $100$~Mpc. All these cases can be understood by noting the different curves of the \Fig\ref{fig:energy_loss_lengths}, as the optimal distance to the source corresponds roughly to the energy loss length of the respective isotope at the highest energy.

In all cases we observe that the spectral data, rather than the composition observables, are the main factor driving the fit, due to their overall lower uncertainties. This means that with current statistics, we cannot point to any particular isotope (from the five mass groups up to iron-56) for the local source that leads to a significantly better joint fit. However, we could conclude from the joint fit for an energy-independent shift, that a local source that emits cosmic rays heavier than nitrogen, within a distance of around 100~Mpc, may be responsible for the differences between the PAO and TA spectrum data.    

\begin{deluxetable}{l|r|r|r|r|r}[htpb!]
    \centering
    \caption{Best-fit parameters from the joint fit to PAO and TA data, assuming the emission of different isotopes from the local source that do not provide the best fit. This complements the information in  \Tab\ref{tab:main_result_parameters}, where we only provide the parameter values for the elements that provide the best fit in each scenario. The best-fit values of the energy-independent systematic shifts are the same in all cases, as listed in \Tab\ref{tab:main_result_parameters}.}
    \label{tab:other_isotopes}
    \tablehead{
        & H
        & He 
        & N
        & Si
        & Fe
    }

    \startdata
    \multicolumn{1}{c|}{\textsc{Sibyll~2.3c}}
    & $\,$
    & $\,$
    & $\,$
    & $\,$
    & $\,$
    \\
    \gammalocal
        &$<-1.0$
        &$<-1.2$
        &$<-1.1$
        &$<-1.0$
        &$<0.4$
    \\
    \Rmaxlocal~(GV)
        &$>8.9\times 10^{10}$
        &$8.9_{-0.3}^{+1.1}\times 10^{9}$
        &$2.2_{-0.3}^{+0.3}\times 10^{9}$
        &$1.3_{-0.1}^{+0.2}\times 10^{9}$
        &$7.9_{-0.1}^{+14.4}\times 10^{8}$
    \\
    \lumlocal~(erg $s^{-1}$)
        &$1.8_{-0.7}^{+0.5}\times 10^{44}$
        &$<9.6\times 10^{38}$
        &$<4.7\times 10^{39}$
        &$<3.7\times 10^{42}$
        &$6.7_{-4.8}^{+32.6}\times 10^{43}$
    \\
    \Dlocal~(Mpc)
        &$176.2_{-46.1}^{+18.7}$
        &$<0.5$
        &$<0.9$
        &$<25.6$
        &$106.2_{-48.5}^{+109.5}$
    \\
     $\chi^2$ / d.o.f.
        & 89.5/50
        & 88.0/50
        & 70.1/50 
        & 67.8/50
        & 69.0/50
    \\
    \multicolumn{1}{r|}{$\left(\chi_{\mathrm{spectrum}} ^\TA \right)^2$}
        & 29.2
        & 18.2
        & 15.3
        & 14.4
        & 13.4
    \\
    \multicolumn{1}{r|}{$\left(\chi_{\langle X_\mathrm{max} \rangle} ^\TA \right)^2$}
        & 4.8
        & 6.9
        & 4.7
        & 4.5
        & 5.1
    \\
    \multicolumn{1}{r|}{$\left(\chi_{\sigma(X_\mathrm{max})} ^\TA \right)^2$}
        & 15.1
        & 22.4
        & 9.6
        & 8.5
        & 10.1
    \\
    \hline
    \multicolumn{1}{c|}{\textsc{Epos-LHC} }
    & $\,$
    & $\,$
    & $\,$
    & $\,$
    & $\,$
    \\
    \gammalocal
        &$<-0.8$
        &$<-1.2$
        &$<-1.1$
        &$<-1.0$
        &$<0.3$
    \\
    \Rmaxlocal~(GV)
        &$>7.1\times 10^{10}$
        &$1.0_{-0.1}^{+0.3}\times 10^{10}$
        &$2.5_{-0.3}^{+0.3}\times 10^{9}$
        &$1.3_{-0.1}^{+0.3}\times 10^{9}$
        &$8.9_{-1.8}^{+13.5}\times 10^{8}$
    \\
    \lumlocal~(erg $s^{-1}$)
        &$7.0_{-6.9}^{+11.0}\times 10^{45}$
        &$<1.2\times 10^{39}$
        &$<6.6\times 10^{39}$
        &$<2.6\times 10^{42}$
        &$6.9_{-5.4}^{+39.5}\times 10^{43}$
    \\
    \Dlocal~(Mpc)
        & $159.2_{-53.1}^{+16.9}$
        &$<0.5$
        &$<1.0$
        &$<18.8$
        &$95.9_{-48.8}^{+119.7}$
    \\
     $\chi^2$ / d.o.f.
        & 102.4/50
        & 101.4/50
        & 91.2/50
        & 91.8/50
        & 92.8/50
    \\
    \multicolumn{1}{r|}{$\left(\chi_{\mathrm{spectrum}} ^\TA \right)^2$}
        & 20.3
        & 16.5
        & 13.7
        & 13.5
        & 12.0
    \\
    \multicolumn{1}{r|}{$\left(\chi_{\langle X_\mathrm{max} \rangle} ^\TA \right)^2$}
        & 6.8
        & 6.0
        & 7.3
        & 8.8
        & 10.3
    \\
    \multicolumn{1}{r|}{$\left(\chi_{\sigma(X_\mathrm{max})} ^\TA \right)^2$}
        & 13.7
        & 17.4
        & 8.7
        & 8.0
        & 9.0
    \\
    \hline
        \multicolumn{1}{c|}{\textsc{QGSJET-II-04}}
    & $\,$
    & $\,$
    & $\,$
    & $\,$
    & $\,$
    \\
    \gammalocal
        &$<0.6$
        &$<-1.3$
        &$<-1.2$
        &$<-1.2$
        &$<-0.1$
    \\
    \Rmaxlocal~(GV)
        &$>6.3\times 10^{10}$
        &$1.1_{-0.1}^{+0.1}\times 10^{10}$
        &$3.2_{-0.3}^{+0.4}\times 10^{9}-$
        &$2.0_{-0.4}^{+0.4}\times 10^{9}$
        &$2.2_{-1.2}^{+0.6}\times 10^{9}$
    \\
    \lumlocal~(erg $s^{-1}$)
        &$2.7_{-2.1}^{+0.6}\times 10^{46}$
        &$<1.2\times 10^{39}$
        &$1.4_{-0.3}^{+7.7}\times 10^{39}$
        &$1.6_{-1.5}^{+2.3}\times 10^{42}$
        &$3.2_{-2.8}^{+1.5}\times 10^{44}$
    \\
    \Dlocal~(Mpc)
        & $194.9_{-51.0}^{+10.2}$
        &$<0.5$
        &$<0.6$
        &$<28.3$
        &$143.9_{-86.2}^{+15.3}$
    \\
     $\chi^2$ / d.o.f.
        & 246.9/50 
        & 257.0/50 
        & 251.1/50
        & 250.4/50
        & 249.5/50
    \\
    \multicolumn{1}{r|}{$\left(\chi_{\mathrm{spectrum}} ^\TA \right)^2$}
        & 27.8
        & 36.7
        & 33.4
        & 31.2
        & 30.1
    \\
    \multicolumn{1}{r|}{$\left(\chi_{\langle X_\mathrm{max} \rangle} ^\TA \right)^2$}
        & 11.7
        & 9.8
        & 12.1
        & 13.9
        & 14.0
    \\
    \multicolumn{1}{r|}{$\left(\chi_{\sigma(X_\mathrm{max})} ^\TA \right)^2$}
        & 6.6
        & 9.8
        & 5.0
        & 5.0
        & 4.8
    \\
    \hline
    \hline
    \enddata
\end{deluxetable}

\clearpage

\section{An ``exotic'' scenario: extreme local accelerator}
\label{app:exotic}

As mentioned in the main text, in addition to our main result, the parameter space of the local source contains another region that can provide a good joint fit to PAO and TA data. This is a scenario where the local source emits cosmic rays with extremely high maximum energies, above $10^{12}$~GeV. For the sake of brevity and simplicity we present only the results of the energy-independent shift. It should be noticed that the results for an energy-dependent shift are very similar.

\begin{figure}[htpb!]
\begin{tabular}{cc}
  \includegraphics[width=.486\textwidth]{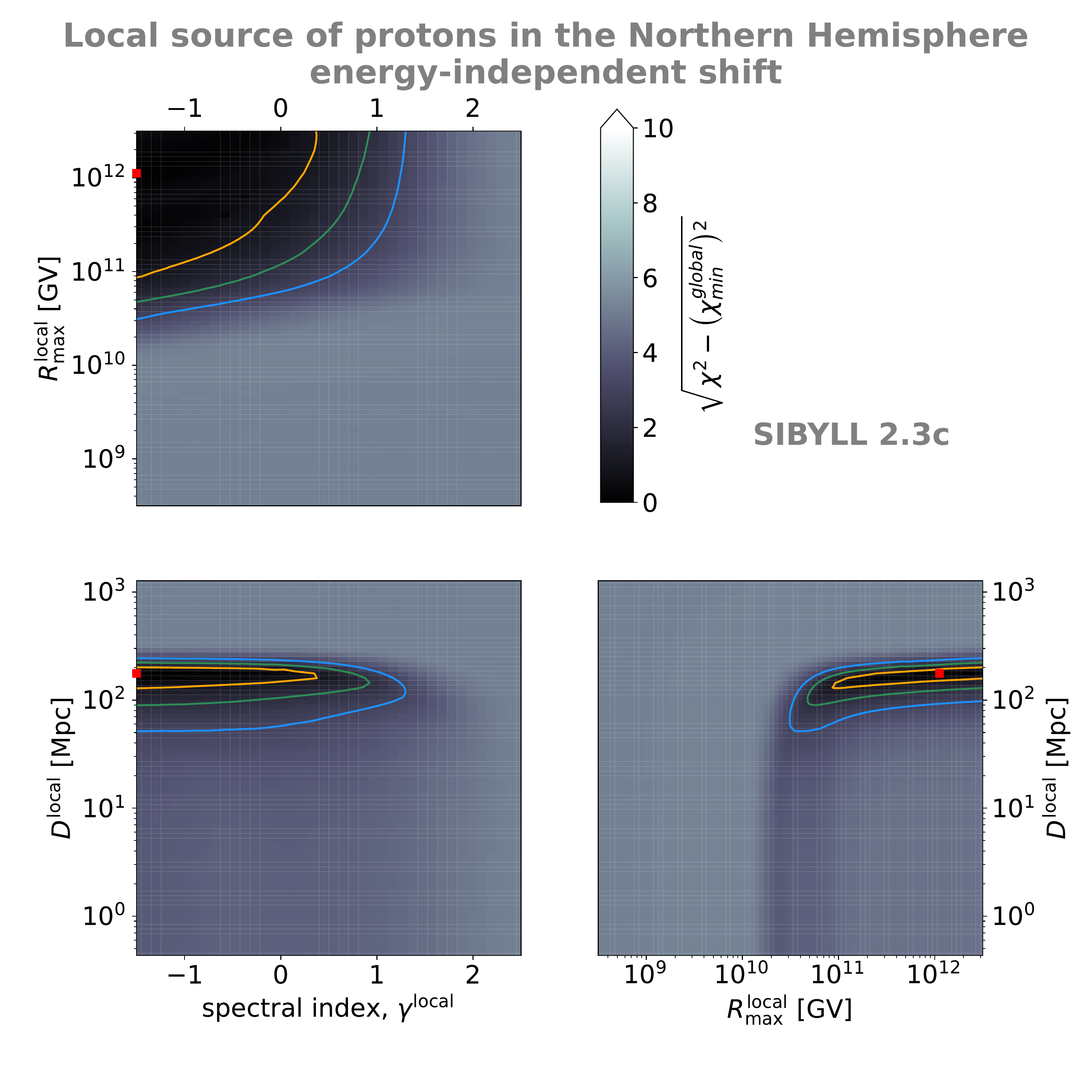}&   \includegraphics[width=.49\textwidth]{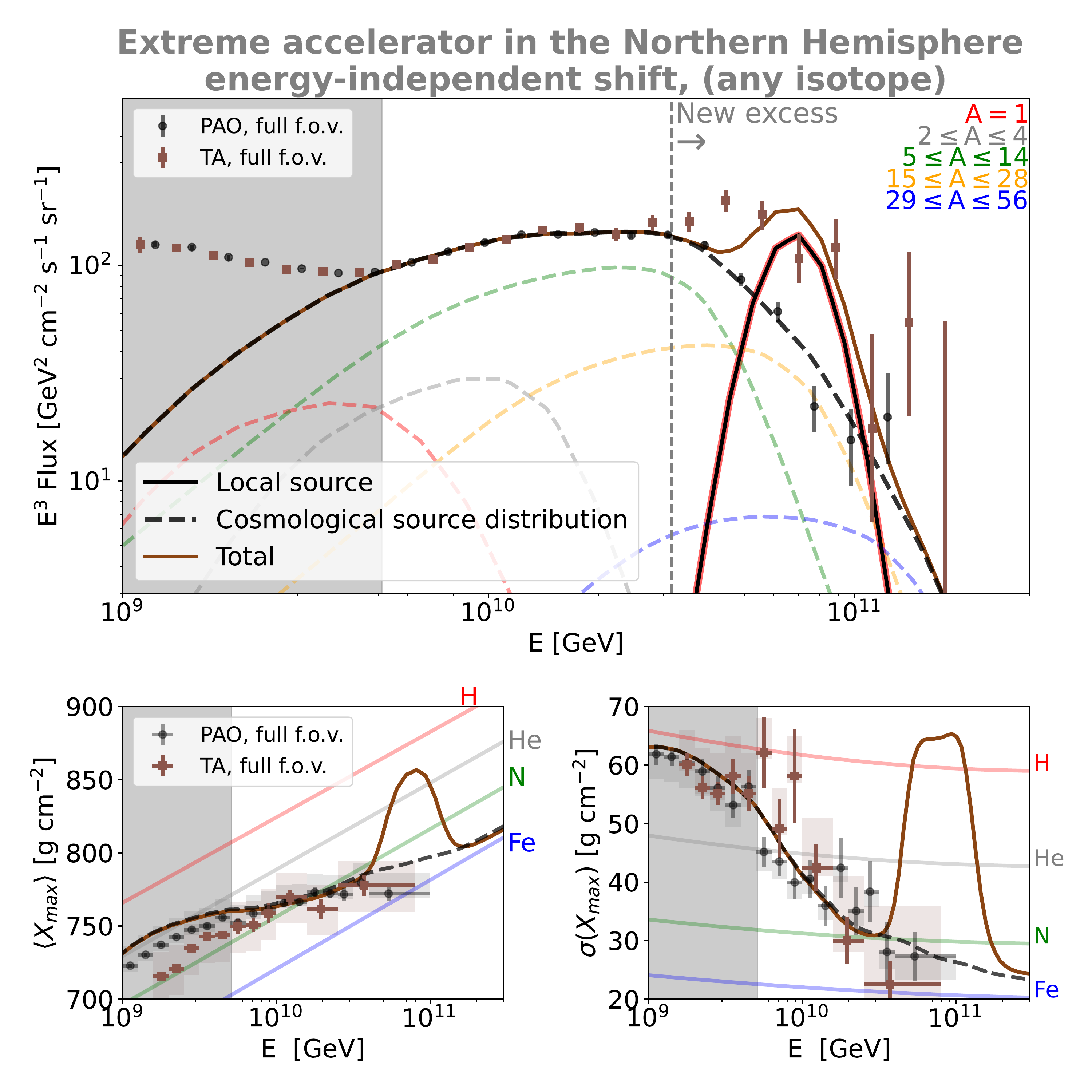} \\
\end{tabular}
\caption{\textit{Left:}  Parameter space of a local source for a joint fit to PAO and TA data when considering a cosmological source distribution and a local source in the Northern Hemisphere that emits a pure protons composition. \textit{Right:}  Spectra (upper panels) and composition observables (lower panels) predicted by the models corresponding to the best-fit case for an extreme accelerator scenario.}
\label{fig:exotic}
\end{figure}

An example of this kind of ``exotic'' solution was represented as red squares in the right-hand panel of \Fig\ref{fig:main_result_contours_local_source}, for the case where the local source emits cosmic rays of the silicon mass group, and considering Sibyll as the air shower model. However, as we demonstrate in this appendix, the extreme source can emit a composition dominated by any mass group. In the left panel of \Fig\ref{fig:exotic} we show as red squares the best-fit parameters of the extreme local source obtained for a pure proton composition and using \textsc{Sibyll 2.3}c as the air shower model. In \Tab\ref{tab:exotic} we provide the complete list of the best-fit parameters of the extreme accelerator for different emitted mass groups and assuming different air shower models. As we can see, the best-fit parameters are close to those obtained in our baseline model (\Fig\ref{fig:main_result_contours_local_source}): a maximum energy of order $\sim$10~ZeV, a distance to the extreme local source of $\sim$100~Mpc, and a hard spectral index. The luminosity of the local source is similar regardless of the emitted isotope, as we can see in \Tab\ref{tab:exotic}. 

To understand these results we turn to the right-hand panel of \Fig\ref{fig:exotic}, where we show the predictions for the spectrum and composition observables. These plots appear equal regardless of the emitted composition. As we can see, the ZeV cosmic rays emitted by the local source suffer strong photodisintegration due to the long distance traveled, to the point where the flux arriving at Earth is completely dominated by protons. These secondary protons carry approximately the same Lorentz factor as the primary nuclei emitted by the source, apart from energy loss processes like pair production and the adiabatic expansion of the Universe. Therefore, as long as the emitted cosmic rays peak at about 10~ZeV, the proton flux observed by TA will peak at a few tens of EeV, thus explaining the excess observed by TA like in the scenarios discussed previously.

Because in this scenario the local source contributes exclusively with protons to the TA spectrum, the result predicts a high value of $\langle X_\mathrm{max}\rangle$ and $\sigma(X_\mathrm{max})$ above a few tens of EeV in TA, as we can see in the bottom-right-hand-side plots of \Fig\ref{fig:exotic}. The only contribution of heavier cosmic rays is then provided by the cosmological source distribution (also observed by PAO). The reason why this scenario can still provide an acceptable fit is that the TA composition observables are not well constrained at these very high energies. However, we must also note that the fit quality provided by an extreme accelerator is overall worse compared to our baseline model, as we can see by the values of $\chi^2$/d.o.f. listed in \Tab\ref{tab:exotic}.

\begin{deluxetable}{l|r|r|r|r|r}[htpb!]
    \centering
    \caption{Best-fit parameter values obtained from a joint fit to PAO and TA data, for the case of an extreme local accelerator in the ZeV regime. We show the results for different emitted mass groups, considering Sybill as the air shower model, and assuming an energy-independent systematic shift in the energy scales of the experiments. The best-fit values of the systematic shifts are the same for all isotopes, as listed in \Tab\ref{tab:main_result_parameters}. We note that the parameters for silicon are depicted as red squares in the left panel of \Fig\ref{fig:main_result_contours_local_source} (baseline model), and the case of protons is shown in the left panel of \Fig\ref{fig:exotic}.}
    \label{tab:exotic}
    \tablehead{
        & H
        & He 
        & N
        & Si
        & Fe
    }

    \startdata
    \multicolumn{1}{c|}{\textsc{Sibyll~2.3c}}
    & $\,$
    & $\,$
    & $\,$
    & $\,$
    & $\,$
    \\
    \gammalocal
        &$<0.3$
        &$<0.0$
        &$<0.0$
        &$<0.4$
        &$<0.1$
    \\
    \Rmaxlocal~(GV)
        &$>1.0\times 10^{11}$
        &$>3.0\times 10^{11}$
        &$>2.2\times 10^{11}$
        &$>1.8\times 10^{11}$
        &$>1.6\times 10^{11}$
    \\
    \lumlocal~(erg $s^{-1}$)
        &$7.6_{-7.4}^{+14.7}\times 10^{45}$
        &$1.8_{-1.7}^{+0.4}\times 10^{45}$
        &$1.3_{-1.3}^{+0.3}\times 10^{45}$
        &$3.5_{-3.2}^{+1.4}\times 10^{45}$
        &$1.7_{-1.2}^{+0.7}\times 10^{44}$
    \\
    \Dlocal~(Mpc)
        & $176.2_{-46.1}^{+18.7}$
        & $176.2_{-46.1}^{+18.7}$
        & $176.2_{-46.1}^{+18.7}$
        &$13.9_{-13.4}^{+9.2}$
        &$95.9_{-43.8}^{+99.0}$
    \\
     $\chi^2$ / d.o.f.
        & 88.3 / 40
        & 88.3 / 40
        & 88.3 / 40
        & 88.3 / 40
        & 88.3 / 40
    \\
    \hline
    \multicolumn{1}{c|}{\textsc{Epos-LHC} }
    & $\,$
    & $\,$
    & $\,$
    & $\,$
    & $\,$
    \\
    \gammalocal
        &$<-0.8$
        &$<-1.2$
        &$<-1.1$
        &$<-1.0$
        &$<0.3$
    \\
    \Rmaxlocal~(GV)
        &$>6.3\times 10^{10}$
        &$8.9_{-0.3}^{+1.1}\times 10^{9}$
        &$2.5_{-0.3}^{+0.3}\times 10^{9}$
        &$1.3_{-0.1}^{+0.3}\times 10^{9}$
        &$8.9_{-1.8}^{+13.5}\times 10^{8}$
    \\
    \lumlocal~(erg $s^{-1}$)
        &$3.0_{-1.8}^{+0.8}\times 10^{44}$
        &$<1.2\times 10^{39}$
        &$<6.6\times 10^{39}$
        &$4.7_{-4.7}^{+20.5}\times 10^{41}$
        &$6.9_{-5.4}^{+39.5}\times 10^{44}$
    \\
    \Dlocal~(Mpc)
        &$143.9_{-26.4}^{+15.3}$
        &$<0.5$
        &$<1.1$
        &$9.2_{-9.2}^{+11.6}$
        &$106.2_{-54.0}^{+109.4}$
    \\
     $\chi^2$ / d.o.f.
        & 100.3 / 40
        & 100.4 / 40
        & 100.3 / 40
        & 100.3 / 40
        & 100.2 / 40
    \\
    \hline
        \multicolumn{1}{c|}{\textsc{QGSJET-II-04}}
    & $\,$
    & $\,$
    & $\,$
    & $\,$
    & $\,$
    \\
    \gammalocal
        &$<-0.6$
        &$<-1.2$
        &$<-1.3$
        &$<-1.2$
        &$<-0.3$
    \\
    \Rmaxlocal~(GV) 
        &$>8.0\times 10^{10}$
        &$>1.6\times 10^{11}$
        &$>1.8\times 10^{10}$
        &$>1.6\times 10^{10}$
        &$>1.4\times 10^{10}$
    \\
    \lumlocal~(erg $s^{-1}$)
        &$2.7_{-2.1}^{+0.6}\times 10^{46}$
        &$1.3_{-0.8}^{+1.4}\times 10^{45}$
        &$9.4_{-4.5}^{+11.1}\times 10^{35}$
        &$3.4_{-0.6}^{+2.5}\times 10^{45}$
        &$7.8_{-1.9}^{+9.7}\times 10^{43}$
    \\
    \Dlocal~(Mpc)
        & $194.9_{-51.0}^{+10.2}$
        &$176.2_{-46.1}^{+18.7}$
        &$176.2_{-46.1}^{+18.7}$
        &$176.2_{-46.1}^{+18.7}$
        &$176.2_{-46.1}^{+18.7}$
    \\
     $\chi^2$ / d.o.f.
        & 252.8/ 40
        & 252.9 / 40
        & 252.8 / 40
        & 252.8 / 40
        & 252.8 / 40
    \\
    \hline
    \hline
    \enddata
\end{deluxetable}

\clearpage
\section{Fitting TA data with only a cosmological source distribution}
\label{app:TA}

\begin{figure}[htpb!]
    \begin{tabular}{cc}
    \includegraphics[width=.486\textwidth]{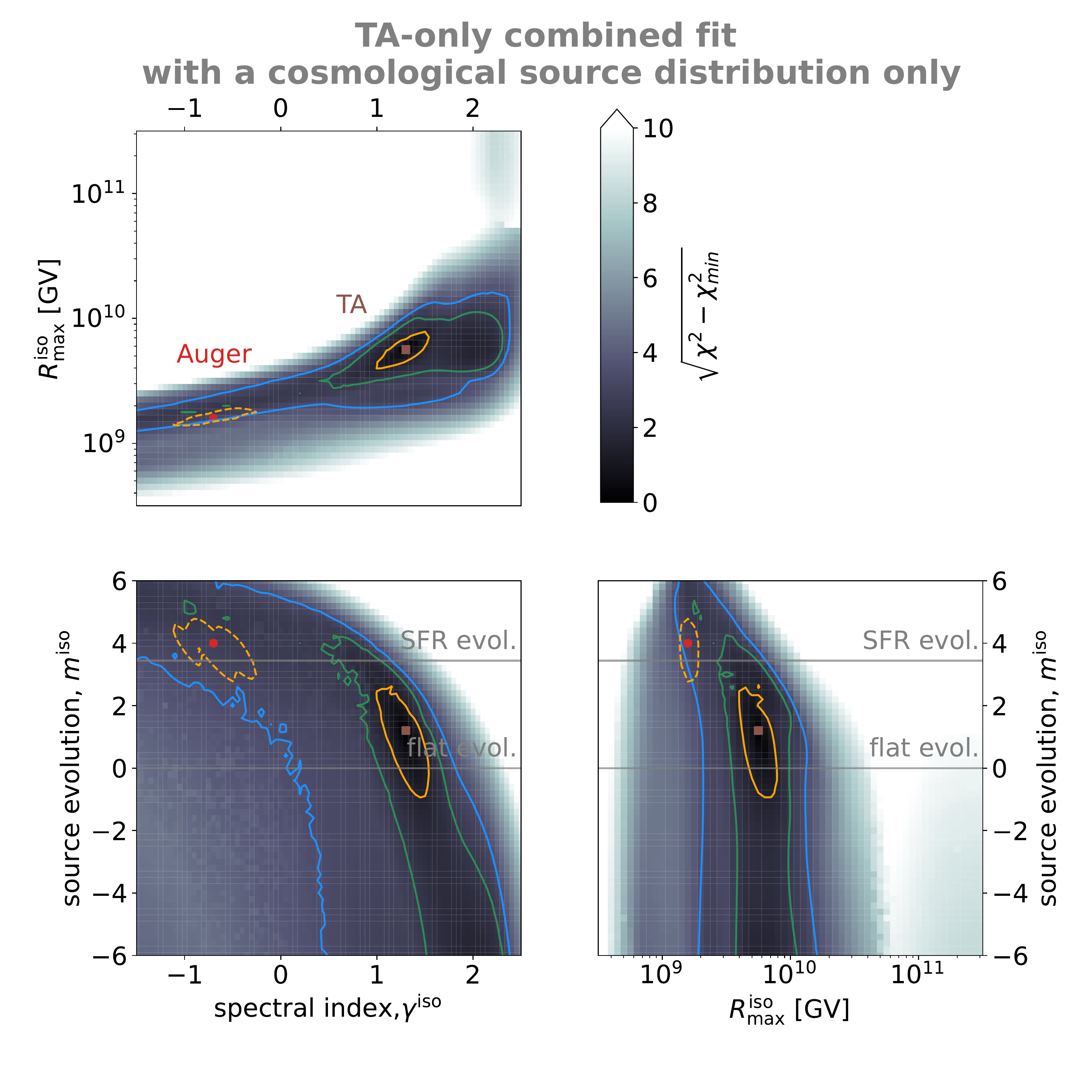} &
    \includegraphics[width=.486\textwidth]{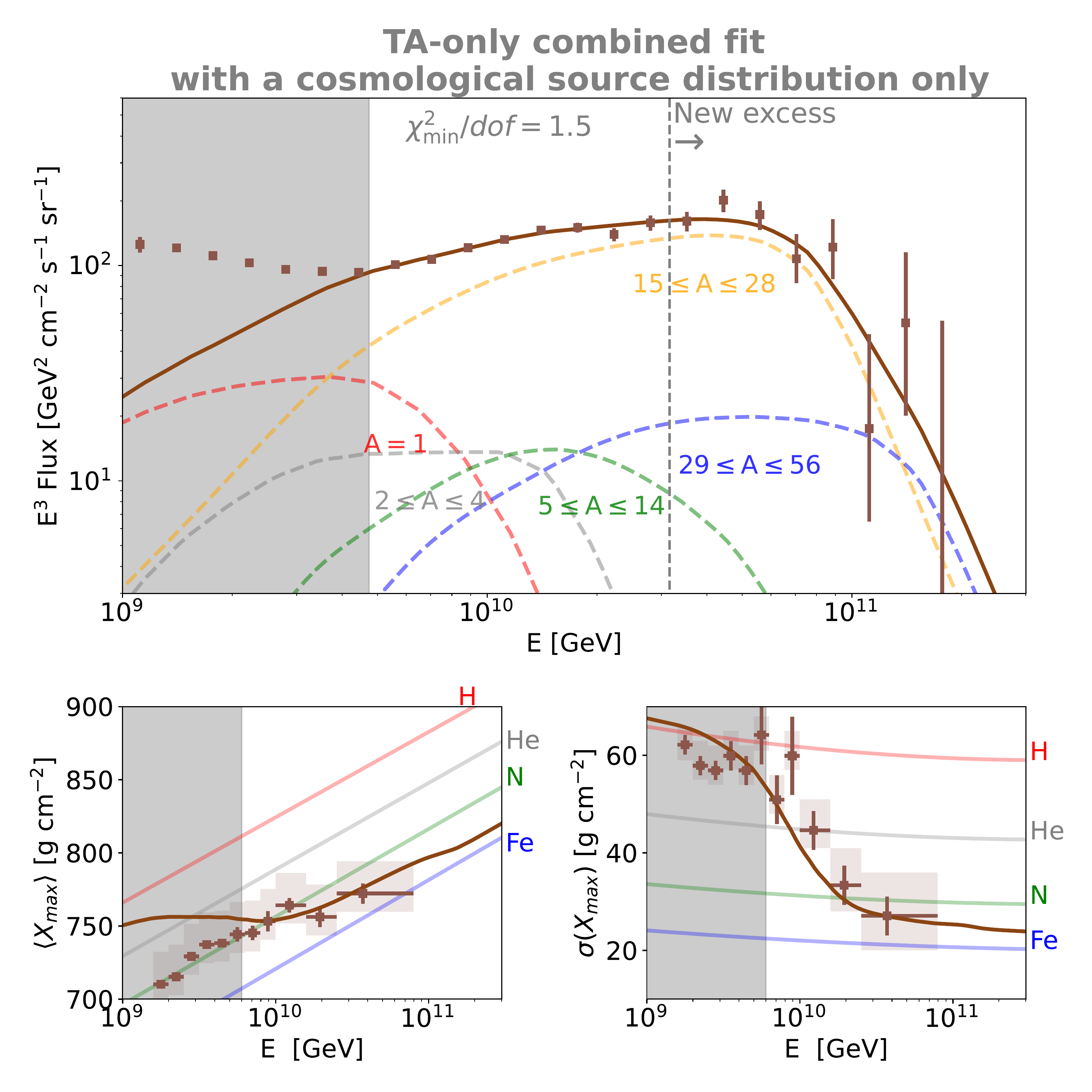}\\

    \end{tabular}
    \caption{\textit{Left:} 
    Parameter space of a cosmological source distribution, based on a fit to the TA spectrum and composition data. The best-fit parameters, listed in \Tab\ref{tab:TA_only_results_parameters}, are shown as a brown dot and for comparison we show as a red dot the best-fit to PAO data~\citep{Heinze:2019}. The solid yellow, green and blue contours correspond to the 1-, 2- and $3\sigma$ regions of our TA-only fit, while the dashed yellow line is the $1\sigma$ region of the PAO-only fit. A direct comparison is not possible because of the different values of $\chi^2_\text{min}$ and numbers of d.o.f. for the two combined fits. \textit{Right:} Spectra (upper panels) and composition observables (lower panels) corresponding to the best-fit parameters of our TA-only fit.}
    \label{fig:TA_only}
\end{figure}

For the sake of comparison we now evaluate how well a cosmological source distribution can describe the TA data set above 5 EeV (i.e. neglecting now PAO measurements). The best-fit parameters of the cosmological source distribution are marked with brown squares in the left panel of  \Fig\ref{fig:TA_only}. For comparison we show as red dots the best-fit parameters obtained by~\citet{Heinze:2019} when fitting only PAO data. On the right-hand panel we show the predicted observables for our TA-only fit. The respective parameter values are listed in \Tab\ref{tab:TA_only}.

As we can see, TA data can be fitted with a cosmological source distribution with a value of $\chi^2/\mathrm{d.o.f.}=22.9/15=1.5$, in the case where Sibyll is considered, while for the other two air shower models that value is higher, as listed in \Tab\ref{tab:TA_only}. These results are similar to the second-best minimum obtained by~\citet{Bergman:2019/7}, who considered \textsc{Epos-LHC} and QGSJET-II-04 as air shower models. We do not obtain the same best-fit minimum obtained by \citet{Bergman:2019/7} because of differences in the analysis method:~firstly, we consider \mcosmo~as a free parameter, while~\citet{Bergman:2019/7} fix its value to \mcosmo$=3$. Secondly, in that work, the authors use the $\xmax$ distributions, while we base ourselves only on the mean and variance values of the  $\xmax$ distribution, which are the only publicly available data. More information would be necessary for a more detailed analysis of the differences between these two results.
    
\begin{deluxetable}{l|r|r|r}[htpb!]
    \centering
    \caption{\label{tab:TA_only_results_parameters}Best-fit parameters obtained from the fit to the TA spectrum and composition data, assuming a cosmological source distribution (and no additional local source). The results are shown for the three different air shower models tested. The case where Sibyll was considered is shown in \Fig\ref{fig:TA_only}.}
    \label{tab:TA_only}
    \tablehead{
        &Sibyll~2.3c
        &\textsc{Epos-LHC}
        &QGSJET-II-04
    }
    \startdata
    \gammacosmo
        & $1.40_{-0.15}^{+0.10}$
        & $1.40_{-0.10}^{+0.15}$
        & $0.10_{-0.30}^{+0.15}$
    \\
    \Rmaxcosmo~(GV)
        & $7.1_{-1.5}^{+0.9}\times10^{9}$
        & $7.9_{-0.9}^{+2.1}\times10^{9}$
        & $3.2_{-0.3}^{+0.3}\times10^{9}$
    \\
    \mcosmo
        & $-0.8_{-2.2}^{+1.0}$
        & $<-4$
        & $<-5.2$
    \\
    $I^9_A (\%)$
        & $\,$
        & $\,$
        & $\,$
    \\
    \multicolumn{1}{r|}{H}
        & $17.6_{-13.5}^{+34.2}$
        & $22.4_{-9.4}^{+13.3}$
        & $5.0_{-4.9}^{+93.3}$
        \\
    \multicolumn{1}{r|}{He}
        & $0.6_{-0.6}^{+99.4}$
        & $0.0_{-0.0}^{+21.7}$
        & $50.4_{-10.2}^{+10.1}$
    \\
    \multicolumn{1}{r|}{N}
        & $1.7_{-1.7}^{+84.2}$
        & $30.7_{-7.4}^{+8.5}$
        & $39.5_{-1.6}^{+1.7}$
    \\
    \multicolumn{1}{r|}{Si}
        & $78.1_{-0.8}^{+0.7}$
        & $46.2_{-5.5}^{+5.6}$
        & $4.0_{-1.1}^{+1.5}$
    \\
    \multicolumn{1}{r|}{Fe}
        & $2.0_{-1.1}^{+2.3}$
        & $0.7_{-0.7}^{+70.5}$
        & $1.0_{-0.3}^{+0.5}$
    \\
    \hline    
    \deltaETA (\%)
        & $-17.6_{-3.4}^{+3.9}$
        & $-14.1_{-3.9}^{+5.4}$
        & $21.0_{-1.5}^{+0.0}$
    \\
    \deltaMeanXmaxTA (\%)
        & 0 (fixed)
        & 0 (fixed)
        & 0 (fixed)
    \\
    \deltaSigmaXmaxTA (\%)
        & 0 (fixed)
        & 0 (fixed)
        & 0 (fixed)
    \\
    \hline
    $\chi^2$ / d.o.f.
        & 22.9 / 15
        & 30.2 / 15
        & 50.9 / 15
    \\
    \hline
    \enddata
\end{deluxetable}

\clearpage

\end{document}